\documentclass[aps,prb,twocolumn,showpacs]{revtex4}
\usepackage{amsmath}
\usepackage{amssymb}
\usepackage{graphicx}

\newcommand{\bvec}[1]{{\mathbf #1}}

\newcommand{\wct}{\omega_c\tau}

\begin{document}

\title{Quantum and classical surface acoustic wave induced magnetoresistance
oscillations in a 2D electron gas}
\author{Malcolm P. Kennett$^{1}$, John P. Robinson$^2$, Nigel R. Cooper$%
^{1}$, and Vladimir I. Fal'ko$^{2,3}$ }
\affiliation{$^{1}$ Cavendish Laboratory, University of Cambridge, Madingley Road,
Cambridge CB3 0HE, UK}
\affiliation{$^{2}$ Physics Department, Lancaster University, LA1 4YB, UK}
\affiliation{$^{3}$ The Abdus Salam ICTP, Strada Costiera 11, I-34014 Trieste, Italy}
\date{\today}

\begin{abstract}
We study theoretically the geometrical and temporal commensurability oscillations induced in the
resistivity of 2D electrons in a perpendicular magnetic field by 
surface acoustic waves (SAWs). We show that there is
a positive anisotropic dynamical classical
contribution and an isotropic non-equilibrium quantum
contribution to the resistivity. We describe how the
commensurability oscillations modulate the resonances in the SAW-induced
resistivity at multiples of the cyclotron frequency. \ We study the effects
of both short-range and long-range disorder on the
resistivity corrections for both the classical and quantum non-equilibrium cases.
We predict that the quantum correction will give rise to
zero-resistance states with associated  geometrical
commensurability oscillations  
at large SAW amplitude for sufficiently large inelastic scattering
times.  These zero resistance states are qualitatively similar to those
observed under microwave illumination, and their nature  
depends crucially on whether the disorder is short- or long-range.
 Finally, we discuss the implications of our
results for current and future experiments on two dimensional electron gases.

\end{abstract}

\maketitle

\section{Introduction}

Recent experiments on high mobility
two-dimensional electron gases (2DEGs)  have shown a variety of 
magnetoresistance effects 
under intense microwave (MW) illumination. 
\cite{Zudov1,Mani1,Zudov2,Mani2,Willett2,Studenikin,Yang,KuKuMPL,Dorozhkin,Du}
One new phenomenon that has attracted considerable attention is MW-induced
magnetoresistance oscillations.  These are spaced according to 
the ratio of the MW frequency to the cyclotron frequency,
showing a {\it temporal commensurability} between the MW field and
the cyclotron motion.
At high MW power the oscillations develop into zero
resistance states (ZRS).\cite{Mani1,Zudov2}
The phenomenological theory of ZRS by Andreev, Aleiner, and Millis \cite{Andreev}
has explained the effect to be a macroscopic manifestation of negative local
resistivity imposed on the 2DEG by energy pumping from the microwave field.
Several microscopic models have been proposed 
\cite{Aleiner,Durst,Mirlin,ZRStheory} to justify the periodic appearance
of negative resistivity as a function of ratio of the MW frequency to 
the cyclotron frequency.  The consensus is that the effect is of a quantum
nature and related to the non-equilibrium occupation of Landau levels in a
2DEG pumped by microwave radiation.  

A separate phenomenon involving {\it geometrical commensurability} oscillations has
been extensively studied in 2D semiconductor structures subjected to
either surface acoustic wave (SAW) 
or static modulations. \cite{SAW,Willett,Weiss,Beenakker,Gerhardts}
Commensurability between the diameter of the cyclotron orbit, $2R_c$ 
and the wavelength $2\pi/q$ of a coherent sound wave is known to 
manifest itself in magneto-oscillations of the dynamical conductivity
$\sigma_{\omega q}$ of metals -- the geometrical resonance
effect.  This was used in studies of three dimensional metals 
to determine the shape of the Fermi surface. \cite{Pippard}
Geometric resonance effects 
have also been observed in 2DEGs in
the attenuation and renormalization of SAW velocity due
to interactions with electrons \cite{Willett,Shilton} and in the drag effect due to SAWs. 
\cite{Shilton}  There has also been some theoretical study of these SAW-induced
effects. \cite{Falko,Efros,Simon,MW,Halperin} 

In all of these studies, the parameter regimes used have been in the low 
frequency ($\omega \to 0$) limit, such that the SAWs are essentially static.
Experiments have recently begun to investigate
the effect of SAWs on magnetoresistance in a 2DEG. \cite{SAW} 
We have studied this effect theoretically, investigating both geometric and
temporal resonances. \cite{Robinson}
For a spatially periodic SAW field, the
commensurability effect can also be viewed as a resonant SAW interaction
with collective excitations of 2D electrons at finite wavenumbers, \cite%
{Willett,Halperin} enabling one to excite modes otherwise forbidden by
Kohn's theorem. \cite{Kohn}  These experimental developments require
new theory for the response of a 2DEG to electric fields with both spatial and temporal
modulation; this work was started in Ref. \onlinecite{Robinson}
and is continued here.

In this paper we study the non-linear dynamical effect in which SAWs
induce changes in the magneto-resistivity of a high quality electron gas in
the regime of classically strong magnetic fields, $\omega _{c}\tau \gg 1$
and high temperatures $k_BT\gg \hbar \omega _{c}$ ($\omega_c = eB/m^*$ 
is the cyclotron frequency, where $B$ is the magnetic field, $m^*$ the electron
effective mass, and $\tau$ is the transport relaxation time).
We have recently shown the existence of SAW-induced magnetoresistance oscillations 
\cite{Robinson} that reflect both temporal and geometrical resonances in the
SAW attenuation.  
There are two competing contributions to the resistivity corrections: one a classical
SAW induced guiding centre drift of cyclotron orbits, the other a quantum contribution
arising from the modulation of the electron density of states.
In Ref. \onlinecite{Robinson}, we restricted the analysis to 2DEGs with 
short-range disorder (which leads to isotropic scattering), so as 
to understand the main qualitative 
features of the oscillations. We also restricted our analysis of the 
quantum corrections to $\omega\lesssim \omega_c$.  In the present work we extend our previous
analysis of the quantum contribution to higher frequencies,
$\omega \gtrsim \omega_c$, and 
address the experimentally
relevant situation of long-range disorder which arises in modulation-doped systems
and leads to small-angle scattering of electrons.  

The classical contribution to the resistivity correction 
originates from the SAW-induced guiding centre drift
of the cyclotron orbits. For a SAW with
frequency $\omega $, and wavenumber $q$, propagating in the $x$ direction
with speed $s=\omega /q$ (we assume $s$ is small compared to the Fermi velocity 
$\mathrm{v}_{F}$), there is an anisotropic increase in the
resistivity $\rho _{xx}$ (at high fields $\omega _{c}\tau \gg 1,$ this is
equivalent to an increase of conductivity in the transverse direction $%
\sigma _{yy}$), which oscillates as a function of inverse magnetic field.
We show that at $\omega
\gtrsim \omega _{c}$ the resistivity change displays resonances at integer multiples
of the cyclotron frequency $\omega \approx N\omega _{c}$.  We find that the
main difference between long- and short-range disorder is that long-range disorder
leads to an effective transport time $\tau^* = 2\tau/(qR_c)^2$ when $qR_c \gg \omega/\omega_c$.  
For small-angle scattering
we obtain analytic formulae for the resistance correction for the limits
$\wct^* \gg 1$ when $(qR_c)^2 \ll \tau/\tau_s$, and $\wct^* \ll 1$  for both 
$(qR_c)^2 \ll \tau/\tau_s$ and $(qR_c)^2 \gg \tau/\tau_s$, where $\tau_s$ is the total
scattering time.\cite{MirlinW}

The quantum contribution arises from the modulation of the electron density
of states (DOS), $\tilde{\gamma}(\epsilon )$, and consequently, from the energy dependence of the
non-equilibrium population of excited electron states caused by Landau level
quantization.
We follow the idea proposed in Ref.~\onlinecite{Aleiner} to explain
the formation of ZRS \cite{Mani1,Zudov2,Andreev} under
microwave irradiation with $\omega \gtrsim \omega _{c}$. We show that in the
frequency range $\tau ^{-1}\lesssim \omega \lesssim \omega _{c}$ the quantum
contribution suppresses resistivity both in $\rho _{xx}$ and $\rho _{yy}$
and persists up to temperatures $k_{B}T\gg \hbar \omega _{c}$ and filling
factors $\nu \gg 1$ where no Shubnikov-de Haas (SdH) oscillations would be seen in
the linear-response conductivity.  We propose a new class of ZRS, in which geometric
commensurability oscillations overlay the ZRS that would be found in the microwave 
($qR_c \to 0$) limit for a short-range potential.  For a long-range potential 
there are ZRS linked to geometric commensurability oscillations, which are
enhanced (by $O( [qR_c]^2)$) over those induced by isotropic scattering.

The paper is organised as follows.  In Sec.~\ref{sec:qualitative} we give
qualitative arguments to determine the form of the classical magnetoresistance
oscillations in the presence of short- and long-range disorder potentials. 
In Sec.~\ref{sec:classicalKE} we obtain these results rigorously using the
classical kinetic equation, and discuss screening of the SAW field.  In 
Sec.~\ref{sec:quantumKE} we give our analysis of the quantum kinetic equation at 
both low and high frequencies.  Finally, in Sec.~\ref{sec:discuss} we discuss 
our results, in particular their implications for experiments.

\section{Qualitative analysis of Classical Magnetoresistance}
\label{sec:qualitative}

It was shown by Beenakker \cite{Beenakker} that the 
magnetoresistance oscillations in a 2DEG in the presence of a spatially
modulated electric field can be understood from a semi-classical 
point of view by considering the 
guiding centre ($\bvec{E}\times \bvec{B}$) drift of cyclotron orbits which lead to 
enhanced diffusion. We apply a similar method below to 
calculate the guiding center drift in the
presence of an electric field with both temporal and spatial oscillations,
for both isotropic scattering (short-range disorder) and small-angle
scattering (long-range disorder).  Note that the following qualitative
analysis gives quantitative results applicable only to the high-field
regime, $\wct \gg 1$.

The {\it dynamical classical} resistivity 
change $\delta ^{c}\rho _{xx}$ can be tracked back to the
SAW induced drift $Y(t)$ (along the $y$-axis) of the guiding center ($X,Y$) of an
electron cyclotron orbit \cite{Beenakker} and the resulting enhancement of
the transverse ($y$-) component of the electron diffusion coefficient, $D_{yy}$,
\begin{equation*}
\dfrac{\delta ^{c}\rho _{xx}}{\rho _{0}}=\dfrac{\delta D_{yy}}{D_0} ,
\end{equation*}%
where $\rho_0 = 2/\gamma e^2 {\mathrm v}_F^2\tau$ is the Drude resistivity
and $D_0 = R_c^2/2\tau$ is the unperturbed diffusion 
coefficient (where the cyclotron radius $R_c = {\mathrm v}_F/\omega_c$).
The drift is caused by an electric field $\tilde{E}_{\omega q}\cos (qx-\omega t)
\hat{\mathbf{x}}$, where $x(t)$ is the position of the particle.
To lowest order in $\tilde{E}_{\omega q}$, the contribution of the SAW to the guiding 
centre drift velocity is
\begin{eqnarray}
\label{eq:ydot}
\dot{Y}(t) & \simeq & \frac{\tilde{E}_{\omega q}}{B} \cos \left[qx(t) - \omega t\right],
\\
\dot{X}(t) & = & 0,
\end{eqnarray}%
where $x(t)$ is the position of the particle {\it neglecting} the effects
of the SAW, but including the effects of disorder scattering.
The change in the electron diffusion coefficient due to the SAW is then
\begin{equation}
\delta D_{yy}  = \int_0^\infty \left<\dot{Y}(t) \dot{Y}(0) \right> dt ,
\end{equation}
where we average over all particle trajectories $x(t)$ in the disorder potential.

\subsection{Short-range disorder: isotropic scattering}
\label{sec:qualitative_isotropic}
For isotropic scattering, the particle performs free cyclotron orbits,
$x(t) = R_c \sin(\omega_c t + \phi_0) + X_0$, up until a scattering event,
after which the subsequent motion in the SAW potential is 
uncorrelated with its preceding motion, provided $qR_c \gg 1$. 
In this case, averaging over the trajectories and scattering events, 
the change in the diffusion coefficient due to the SAW is
\begin{equation}
\delta D_{yy} =  \int_{0}^{\infty }\, dt \, e^{-t/\tau}
\int_{0}^{2\pi }\int_{0}^{2\pi }\dfrac{d\phi_0}{2\pi }\dfrac{d\psi }{2\pi } \,
\dot{Y}(t) \dot{Y}(0),
\end{equation}%
where $\psi = qX_0$, and $\dot{Y}(t)$ is calculated from Eq.~(\ref{eq:ydot}) 
using the free cyclotron motion. 
Using the Bessel function identity
\begin{equation}
\label{eq:besselid}
e^{iz\sin\phi} = \sum_{N=-\infty}^\infty J_N(z) e^{iN\phi},
\end{equation}
one can obtain the frequency and wave number dependence of the
magnetoresistance effect:
\begin{equation}
\frac{\delta ^{c}\rho _{xx}}{\rho _0}=\frac{\delta D_{yy}}{D_0}=\frac{1}{4}\left( ql\tilde{\mathcal{E}}\right)
^{2}\sum_{N=-\infty }^{\infty }\frac{J_{N}^{2}(qR_{c})}{1+\left( \omega
-N\omega _{c}\right) ^{2}\tau ^{2}},  \label{Fullresult2}
\end{equation}%
where 
\begin{equation}
\tilde{\mathcal{E}} = ea_{\mathrm{scr}}\tilde{E}_{\omega q}^{\mathrm{SAW}}/\epsilon _{F},
\label{eq:sawamp}
\end{equation}
is the dimensionless SAW amplitude.  The electric field $\tilde{E}_{\omega q}$ discussed in this section is a travelling
wave, rather than the standing wave situation discussed in Secs.~\ref{sec:classicalKE}, \ref{sec:quantumKE} 
and \ref{sec:discuss}.  It is
related to the $E_{\omega q}$ discussed in Sec.~\ref{sec:classicalKE} by $\tilde{E}_{\omega q} = 2E_{\omega q}$,
implying $\tilde{\mathcal{E}} = 2\mathcal{E}$.  If we include another travelling wave to
generate a standing wave, then the result in Eq.~(\ref{Fullresult2}) should be multiplied by a factor of 2, and
this reproduces Eq.~(\ref{WithK}).
In Eqs.~(\ref{Fullresult2}) and (\ref{eq:sawamp}),
$\tilde{E}_{\omega q}^{\mathrm{SAW}}$ is the SAW longitudinal electric field, 
$a_{\mathrm{scr}}=\chi /2\pi e^{2}\gamma $ the 2D screening radius, $\epsilon
_{F}$ the Fermi energy, $l=\mathrm{v}_{F}\tau $ the mean free path, $\chi$  the background
dielectric constant, and $\gamma =m/\pi
\hbar^{2}$ the electron density of states. Equation~(\ref{Fullresult2}) includes the
Thomas-Fermi screening of the SAW field by 2D electrons, 
$\tilde{E}_{\omega q}=qa_{\mathrm{scr}}\tilde{E}_{\mathbf{\omega }q}^{\mathrm{SAW}}$, which we discuss in detail in 
Sec.~\ref{sec:screening}.  At large $qR_c$ we can 
expand the Bessel functions to get

\begin{equation}
\frac{\delta ^{c}\rho _{xx}}{\rho _0} \simeq \frac{1}{\pi qR_c}\left( ql\tilde{\mathcal{E}}\right)
^{2}\sum_{N=-\infty }^{\infty }\frac{\cos^2\left(qR_c + 
\frac{N\pi}{2} - \frac{\pi}{4}\right)}{1+\left( \omega
-N\omega _{c}\right) ^{2}\tau ^{2}} . \label{Fullresult3}
\end{equation}

\begin{figure}[h]
\center{\includegraphics[width=8.3cm]{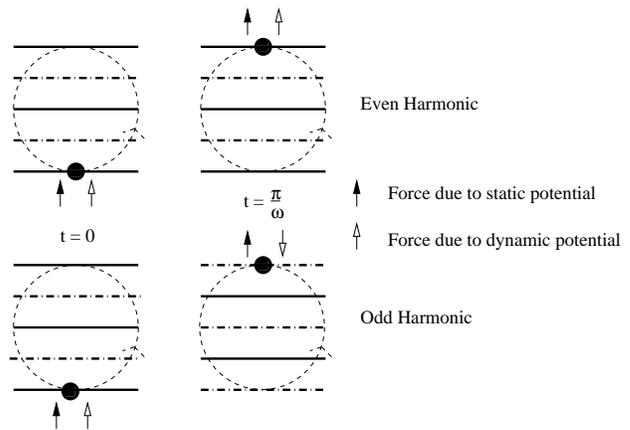}}
\caption{Comparison of electron motion in a static and a dynamic periodic potential.
When the forces on opposite sides of the orbit are in the same direction,
there is a maximum in the magnetoresistance.  The horizontal lines 
correspond to maximum positive (solid lines) and maximum negative (dot-dashed lines) values of the field.
In the dynamic case the direction of the field oscillates with frequency $\omega$.
Thus, when $\omega$ is an odd multiple of the cyclotron frequency $\omega_c$, 
there is an interchange of the resistance maxima and minima
as compared to a static potential.}
\label{fig:qual}
\end{figure}

From Eqs.~(\ref{Fullresult2}) and (\ref{Fullresult3}) we can see that
there are a sequence of resonances at integer multiples of the cyclotron
frequency, $\omega \approx N\omega _{c}$.  The widths of these resonances
are controlled by $\omega\tau$, with large values of $\omega\tau$ leading to
very narrow resonances. The oscillations for even
harmonics are in phase with the Weiss oscillations of the static potential,
whilst those of odd harmonics are $\pi$ out of phase. This can be understood by noting that the
main contributions to the drift occur when the electrons are moving parallel
to equipotential lines. For odd harmonics the phase of
the potential at the half-orbit point is opposite to that for a static
potential, and hence the cancellation and reinforcement effects that lead to
minima and maxima in the resistance are interchanged between the static
and dynamic cases.  This is illustrated via
a comparison of the two situations in Fig. \ref{fig:qual}.  Alternatively,
this can be seen from examination of Eq.~(\ref{Fullresult3}), since the
static case ($\omega = 0$) is dominated by the $N=0$ term in the sum, whereas
all $N$ contribute in the dynamic case.

\begin{figure}[h]
\center{\includegraphics[width=6cm,angle=270]{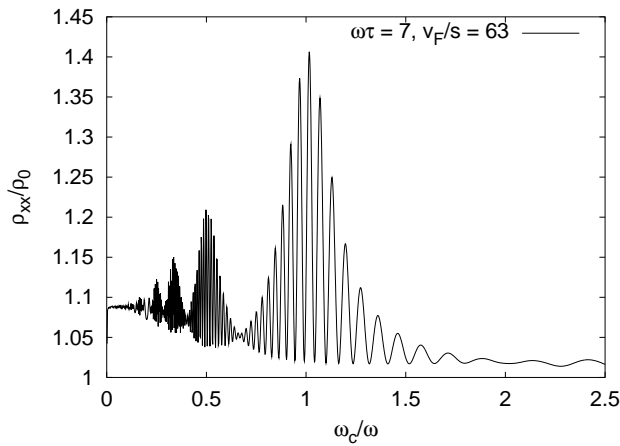}}
\caption{Magnetoresistance for a short-range potential
using the parameters of Ref. 1, and assuming $\omega = 2\pi\times 10$ GHz ($\omega\tau = 7$) 
and ${\mathrm v}_F/s = 63$. We choose a dimensionless SAW amplitude of
 ${\mathcal E} = 0.01$. }
\label{fig:zudov1}
\end{figure}

\begin{figure}[h]
\center{\includegraphics[width=6cm,angle=270]{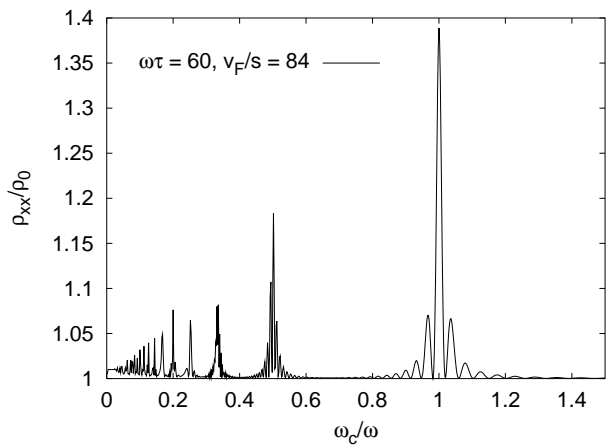}}
\caption{Magnetoresistance for a short-range potential
using the parameters of Ref.~2, and assuming $\omega = 2\pi \times 10$ GHz, which 
implies $\omega\tau = 60$ and ${\mathrm v}_F/s = 84$. We choose ${\mathcal E} = 0.001$, a factor of
10 less than in Fig. 2.}
\label{fig:zudov2}
\end{figure}

In the regime $\omega
_{c}\tau \gg {\mathrm v}_F/s$, the resonances are very narrow and appear to display a
random sequence of heights, rather than the linear in $B$ dependence evident 
in Figs.~\ref{fig:zudov1} and \ref{fig:zudov2}, reflecting the dependence on the geometric
resonance conditions.   
In the intermediate frequency domain, $\tau ^{-1}\ll \omega \ll \omega _{c}$
, a natural regime for GaAs structures with densities $n_{e}\gtrsim 10^{10}
\mathrm{cm}^{-2}$ at sufficiently high magnetic fields, the classical
oscillations take the form 
\begin{equation}
\frac{\delta ^{c}\rho _{xx}}{\rho_0}\approx \frac{\mathrm{v}_{F}^{2}}{4
s^{2}}\tilde{\mathcal{E}}^{2}J_{0}^{2}(qR_{c}).  \label{saturation}
\end{equation}%
The competition between electron screening effects ($\propto q^{2}$),
the dynamical suppression of commensurability by the SAW motion ($\propto
\omega ^{-2}$), and the relation $\omega /q=s$, means that the form of these
oscillations is independent of the absolute value of the SAW frequency,
provided that conditions $\tau ^{-1}\lesssim\omega $ and ${\mathrm v}_F \gg s$ are
satisfied.  Note however, that for a fixed SAW amplitude,  there is still $q$ dependence
of the SAW field $\tilde{E}^{\rm SAW}_{\omega q}$ since the electric field
induced in the 2DEG by the piezoelectric coupling is a
function of $qd$, where $d$ is the distance between the surface and the 2DEG. \cite{Simon2}

We illustrate the interplay between dynamical resonances in the time and
space domains by plotting
Eq.~(\ref{Fullresult2}) as a function of $\omega_c/\omega$ 
for the following experimentally relevant parameters.  In GaAs 
based 2DEGs, $m^* = 0.068 \, m_e$, (where $m_e$ is the electron mass) and 
$s \simeq 3000 \, {\rm ms^{-1}}$.  The highest SAW frequencies that have been 
used in experiments on 2DEGs are $\sim 10$ GHz; \cite{Willett3} we consider this
fequency with the sample densities and mobilities reported in Refs.~\onlinecite{Zudov1,Zudov2}.
In Ref.~\onlinecite{Zudov1}, the mobility $\mu$ is $3 \times 10^6 \, {\rm cm^2/Vs}$, and
the density $n_e = 2 \times 10^{11} \, {\rm cm^{-2}}$, corresponding to $\omega\tau = 7$
and ${\mathrm v}_F/s = 63$ (note that ${\mathrm v}_F/s = qR_c$ when $\omega = \omega_c$).
The magnetoresistance for these parameters is plotted in Fig.~\ref{fig:zudov1}
In Ref.~\onlinecite{Zudov2}, $\mu = 2.5 \times 10^7 \, {\rm cm^2/Vs}$ and $n_e = 3.5 \times 
10^{11} \, {\rm cm^{-2}}$, making it one of the highest mobility samples yet fabricated.
For these parameter values, $\omega\tau = 60$, and ${\mathrm v}_F/s = 84$, and we find a magnetoresistance
trace as shown in Fig.~\ref{fig:zudov2}.  At lower frequencies in such a high quality 2DEG, 
we expect the width of the resonances to broaden and become similar to those shown in 
Fig.~\ref{fig:zudov1}. 

The dynamical mechanism just described dominates in a classical electron
gas. No redistribution of electron kinetic energy (due to SAW absorption)
will additionally change the magnetoresistance until the 2DEG is heated to a
temperature $k_{B}T_{e}\gtrsim \hbar \omega _{c}k_{F}/q$ where geometrical
oscillations become smeared. The essential assumption leading to this
statement is that electron single-particle parameters (velocity, ${\mathrm v}$, and $\tau
^{-1}$) vary slowly with energy at the scales comparable to the Fermi energy
and can thus be approximated by constants.  However, for high-quality 2DEGs (as formed in 
modulation doped GaAs devices) the typical disorder
potential is not simply short-range (as assumed for the isotropic
scattering model considered above), but is dominated by a long-range part due to 
the coulombic potentials associated with remote dopants; this long-range disorder leads to small-angle
scattering.  We now turn to consider this situation.

\subsection{Long-range random potential -- small angle scattering}
\label{sec:qual-sma}

To calculate the resistivity correction in the presence of long range
disorder, we consider an electron undergoing cyclotron motion and
subject to random (small-angle) changes in direction. We express the
position of the particle as
\begin{equation}
\label{eq:position}
 \bvec{r}(t) = \bvec{R}(t) + R_c(\sin\phi(t),\cos\phi(t)) ,
\end{equation}
where $\bvec{R}(t)$ is the position of the guiding centre and $\phi(t)$
is the angular position around the cyclotron orbit. In the absence of
disorder, $\phi(t) = \phi_0 + \omega_c t$ and $\bvec{R}(t)$ is constant
in time. The effect of a disorder potential on $\phi(t)$ may be
represented by random changes: the scattering events are separated by
a characteristic time $\tau_s$ (sometimes referred to as the {\it
quantum lifetime}), and at each scattering event $\phi$ jumps through
a (random) angle of magnitude $\delta\phi$, which is related to the
lengthscale of the disorder potential $L$ by $\delta\phi \sim 1/(k_F
L)$ where $k_F$ is the Fermi momentum.  At any scattering event only
the direction of motion changes; the position $\bvec{r}(t)$ is constant,
so Eq.~(\ref{eq:position}) implies that for an 
instantaneous change $\delta\phi$ in $\phi$, there is a change in guiding centre position of

\begin{equation}
\label{eq:deltar}
\delta\bvec{R} = R_c(-\cos\phi,\sin\phi)\delta\phi .
\end{equation}

In this section we consider the limit of a long range potential, for
which $\delta\phi\ll 1$. In this case the scattering of $\phi$ can be
viewed as leading to a continuous diffusive motion with a diffusion
constant (in angle) of order $\sim (\delta\phi)^2/\tau_s$.  Similarly,
we can view the guiding centre co-ordinates as undergoing continuous
diffusive motion, provided the typical jump [Eq.~(\ref{eq:deltar})] in
guiding centre co-ordinate is small compared to other relevant
lengthscales, in particular the wavelength of the SAW, $R_c \delta\phi
\ll 1/q$. The diffusion constant for the guiding centre is $ \sim
R_c^2 (\delta\phi)^2/\tau_s$.  Note that the diffusion constants for
$\phi$ and $\bvec{R}$ depend only on the {\it transport relaxation time}
$\tau \sim \tau_s/(\delta\phi)^2$, so only $\tau$ enters the
theory. In terms of $\tau$, the conditions for validity of this theory
($\delta\phi\ll 1$, $R_c\delta\phi \ll 1/q$) can be written $\tau/\tau_s\gg
1$ and $(qR_c)^2\ll \tau/\tau_s$.

To model the continuous diffusion of $\phi$ and $\bvec{R}$, we introduce
two sources of Gaussian noise, ${\mathcal L}_1$ and ${\mathcal L}_2$,
with
\begin{equation}
\left<{\mathcal L}_{i}(t^\prime){\mathcal L}_{i}(t)\right> = \Gamma_{i}
\delta(t - t^\prime) ,
\end{equation}
(where $i=1,2$) and write the effects of scattering on the phase of the
orbit and on the $x$-component of the guiding center position as
\begin{eqnarray}
\phi(t) &  = & \phi_0 + \omega_c t + \int^t_0 {\mathcal L}_1(t^\prime) \,
dt^\prime   ,\\
X(t) & = &  X_0 + R_c \int_0^t {\mathcal L}_2(t^\prime) \, dt^\prime .
\end{eqnarray}
The two sources of noise are related through Eq.~(\ref{eq:deltar})
$$\mathcal{L}_2(t) = -R_c\mathcal{L}_1(t) \cos\phi(t) , $$ which requires
that $\Gamma_1 = 2\Gamma_2$ in the situation  of interest where we
average over all trajectories (hence over $\phi_0$). From a direct
evaluation of the classical Kubo formula for the diffusion constant
for this model in the absence of a SAW potential, we find that
\begin{equation}
\label{eq:gammas}
\Gamma_1 = 2\Gamma_2 = \frac{2}{\tau} ,
\end{equation}
where $\tau$ is the conventional definition of the transport
relaxation time, in terms of which the Drude expression for diffusion
constant is $D_{xx}=D_{yy} = \frac{1}{2} v_F^2 \tau/(1+\omega_c^2\tau^2)$.

We simplify subsequent calculations by ignoring correlations of
${\mathcal L}_1$ and ${\mathcal L}_2$ beyond those in Eq.~(\ref{eq:gammas}), which
we expect to be accurate for $\omega_c\tau^* \gg 1$, (with $\tau^*$ defined in 
Eq.~(\ref{eq:taustar}) below) in which case the
particle is able to explore all values of $\phi(t)$ on the timescale
of the relevant scattering time.  We can then calculate the change in the
diffusion coefficient from the guiding centre drift as in
Sec.~\ref{sec:qualitative_isotropic} for isotropic scattering,
averaging over the electron trajectories
\begin{eqnarray}
\delta D_{yy} & = & \int_0^\infty dt \left<\left< \dot{Y}(t)\dot{Y}(0)
\right>\right>_{{\mathcal L}_1,{\mathcal L}_2}
 \nonumber \\
& \simeq & \frac{\tilde{E}_{\omega q}^2}{B^2} {\rm Re}\left\{ \int_0^\infty dt
\int_0^{2\pi} \frac{d\phi_0}{2\pi}
\left<\left< e^{iqR_c(\sin\phi(t) - \sin\phi_0)} \right. \right.
\right. \nonumber \\
& & \left. \left. \left. \hspace*{1cm} \times \,
e^{iqR_c \int^t {\mathcal L}_2(t^\prime) dt^\prime
- i\omega t}\right>\right>_{{\mathcal L}_1,{\mathcal L}_2}\right\} ,
\end{eqnarray}
where our assumption that ${\mathcal L}_1$ and ${\mathcal L}_2$ are
uncorrelated allows us
to perform  the average over noise:
\begin{eqnarray}
\delta D_{yy} &
 \propto & \tilde{E}_{\omega q}^2 {\rm Re}\left\{ \int_0^\infty dt \, e^{-i\omega
t}
 \int_0^{2\pi} \frac{d\phi_0}{2\pi} \right.  \\
 & & \left. \times
\left< e^{iqR_c(\sin\phi(t) - \sin\phi_0)} \right>_{{\mathcal L}_1} \left<
e^{
iqR_c \int^t {\mathcal L}_2(t^\prime) dt^\prime}\right>_{{\mathcal
L}_2}\right\}. \nonumber
\end{eqnarray}
Using
\begin{eqnarray}
\left< e^{iqR_c(\sin\phi(t) - \sin\phi_0)} \right>_{{\mathcal L}_1}
& = & \sum_{N,M = -\infty}^\infty J_N(qR_c)J_M(qR_c) \nonumber \\
& & \hspace*{0.4cm}\times \, e^{i(N-M)\phi_0}e^{iN\omega_c t}
e^{-\frac{N^2\Gamma_1}{2} t}, \nonumber \\
\left< e^{
iqR_c \int^t {\mathcal L}_2(t^\prime) dt^\prime}\right>_{{\mathcal L}_2}
& = & e^{-\frac{(qR_c)^2}{2}\Gamma_2 t}, \nonumber
\end{eqnarray}
we find a resistance correction of
\begin{eqnarray}
\frac{\delta^c\rho_{xx}}{\rho_0} = \frac{{\mathrm v}_F^2 \omega^2
\tau
}{4s^2}
{\tilde{\mathcal E}}^2 \sum_{N=-\infty}^\infty \frac{J_N(qR_c)^2 \,\tau_{N q}}{1 +
(N\omega_c - \omega)^2 \tau_{Nq}^2},
\label{eq:longrangequal}
\end{eqnarray}
where
\begin{equation}
\label{eq:taupq}
\tau_{Nq} = \frac{\tau}{N^2 + \frac{(qR_c)^2}{2}}.
\end{equation}
 The form of this correction
is that found for isotropic scattering, but with $\tau_{Nq}$ replacing
$\tau$
in the correction to the diffusion coefficient. \cite{footnote1}  The result in 
Eq.~(\ref{eq:longrangequal}) can be related to the
result found in Eq.~(\ref{eq:Fullresult-longr}) in a similar manner 
to the case of short-range scattering.

In the experimentally relevant range of parameters, $\omega/\omega_c \ll qR_c$, and 
noting that in the vicinity of resonance $N$, $\omega/\omega_c \simeq N$, we define a scattering time
$\tau^* \ll \tau$ as

\begin{equation}
\frac{1}{\tau^*} \equiv \frac{q^2 R_c^2}{2\tau} ,
\label{eq:taustar}
\end{equation}
and $\tau_{Nq} \simeq \tau^*$ in our regime of interest.
This quantity 
enters in our discussion of the kinetic equation for small-angle scattering in 
Sec.~\ref{sec:long-range-potential}.  Note that the use of this result is also
subject to the constraint $\wct \gg 1$ with the additional requirement
$qR_c^2 \ll l$, which is discussed in more detail in Sec.~\ref{sec:classicalKE}.

\section{Classical Kinetic Equation}
\label{sec:classicalKE}

We analyze the classical kinetic equation for a
2DEG at temperature $k_{B}T\lesssim \hbar \omega _{c}k_{F}/q$ irradiated by
SAWs. Our approach is to solve the kinetic equation for the electron distribution function
\begin{equation}
f(t,x,\varphi ,\epsilon )=f_{T}+\sum\limits_{\omega q}e^{-i\omega
t+iqx}\sum_{m}f_{\omega q}^{m}(\epsilon )e^{im\varphi },
\end{equation}%
using the method of successive approximations. Here, $f_{T}(\epsilon )$ is
the homogeneous equilibrium Fermi function, and the angle $\varphi $ and
kinetic energy $\epsilon $ parametrize the electron state in momentum space.
Each component $f^{m}$ describes the $m$-th angular harmonic of the time-
and space-dependent non-equilibrium distribution. To describe local values
of the electron current and the accumulated charge density, we use the
energy-integrated functions, $g_{\omega q}^{m}=\int_{0}^{\infty }d\epsilon
f_{\omega q}^{m}$. The relaxation of the local non-equilibrium distribution towards a Fermi
function characterized by the value of local Fermi energy, $\epsilon
_{F}(t,x)$ (determined by the local electron density $n(t,x)\propto
g^{0}(t,x)=\int_{0}^{\infty }d\epsilon \, f^{0}(t,x)$, where $f^{0}(t,x)={\displaystyle \int }
\dfrac{d\varphi }{2\pi }f(t,x)$) and the kinetic equation is 

\begin{equation}
\mathcal{\hat{L}}f= \mathcal{\hat{C}}f, 
\label{KinEq}
\end{equation}%
where
\begin{equation}
\mathcal{\hat{L}}=\partial _{t}+\mathrm{v}\cos \varphi \, \partial _{x}+\left[
\omega _{c}-\frac{eE}{p}\sin \varphi \right] \partial _{\varphi }+e\mathrm{v}%
E\cos \varphi \, \partial_{\epsilon} ,
\label{KinEq2}
\end{equation}
with $\hat{\mathcal C}$ the collision integral, $E$ the electric field 
and $p$ the electron momentum.

\subsection{Isotropic scattering} 
\label{sec:short-range-potential}
In the presence of a short-range potential, the scattering
is isotropic and we can use the relaxation time approximation
for the collision integral, which is  
\begin{equation}
\mathcal{\hat{C}}f=-\dfrac{f-f^{0}}{\tau }-\dfrac{f^{0}-f_{T}(\epsilon
-\epsilon _{F}(t,x))}{\tau _{in}} ,  \label{KinEqfull}
\end{equation}%
where we distinguish between the elastic scattering rate $\tau ^{-1}$ and
energy relaxation rate $\tau _{in}^{-1}$. 

The dynamical perturbation of the distribution function can be found from
time/space Fourier harmonics of Eq. (\ref{KinEq}) at the frequency/wave
number of the SAW, 
\begin{equation}
\left[ \partial _{\varphi }+\frac{1}{\omega _{c}\tau }-i\frac{\omega }{%
\omega _{c}}+iqR_{c}\cos \varphi \right]f_{\omega q}=\Psi (\varphi ),
\label{KinHarm}
\end{equation}%
\begin{eqnarray*}
\Psi(\varphi)  &=&-\frac{g_{\omega q}^{0}}{\omega _{c}\tau _{in}}(\partial _{\epsilon
}f_{T})+\dfrac{\tau ^{-1}-\tau _{in}^{-1}}{\omega _{c}}f_{\omega q}^{0} \\
&&-\frac{eE_{\omega q}}{\omega _{c}}[\mathrm{v}\cos \varphi \, \partial
_{\epsilon }-\frac{\sin \varphi }{p}\partial _{\varphi }](f_{00}+f_{T}),
\end{eqnarray*}%
where we include the unknown perturbation of the time/space averaged
function, $f_{00}$, related to the DC current to lowest order in $E_{\omega
q}$. We note that as $\tau_{in}$ only contributes to heating of the 
2DEG, we can ignore it here for the AC part of the distribution 
function, but must retain it for analysis of the DC part of the 
distribution function.  In our expression for $\Psi$ we neglect 
contributions from higher Fourier harmonics of $f$, such as $f_{2\omega\, 2q}$,
$f_{2\omega\, 0}$, and $f_{0\, 2q}$, since these only affect DC transport at
quartic order in the SAW field, whereas our interest is only in effects that 
are quadratic in the SAW field (i.e. linear in the SAW power).  We also
omit terms involving $E_{00} f_{\omega q}$, since they contribute to DC transport
only beyond the linear response regime in the DC field.
Equation (\ref{KinHarm}) can be formally solved using the Green
function $G(\varphi ,\tilde{\varphi})$ 
\begin{eqnarray}
f_{\omega q}(\varphi ) &=&\int_{-\infty }^{\varphi }G(\varphi ,\tilde{\varphi%
})\Psi (\tilde{\varphi})d\tilde{\varphi},  \label{GF1} \\
G(\varphi ,\tilde{\varphi}) &=&e^{[\frac{1}{\omega _{c}\tau }-\frac{i\omega 
}{\omega _{c}}](\tilde{\varphi}-\varphi )+iqR_{c}[\sin \tilde{\varphi}-\sin
\varphi ]}\,,  \label{GF2}
\end{eqnarray}%
which allows for an infinite range of variation of $\varphi $ whilst
guaranteeing periodicity of the solution $f_{\omega q}(\varphi )$.  We
also introduce the useful quantity

\begin{eqnarray}
K &=&\int_{0}^{2\pi }\dfrac{d\varphi }{2\pi }\int_{-\infty
}^{\varphi } \dfrac{d\tilde{\varphi}}{\omega _{c}\tau }G(\varphi
,\tilde{\varphi})  \notag
\\
&=&\sum_{N=-\infty }^{\infty }\dfrac{J_{N}^{2}(qR_{c})}{1+i\tau
(N\omega_{c}-\omega )}\quad  \label{K-1} ,
\end{eqnarray}
which has the properties $$K(\omega ,q)=K^{\ast
}(-\omega ,q)=K(\omega ,-q),$$ and is obtained from
Eqs.~(\ref{GF1}) and (\ref{GF2}) using Eq.~(\ref{eq:besselid}).

\subsubsection{Screening and Dispersive Resonance Shift}

\label{sec:screening}

The electric field $E$ in Eq.~(\ref{KinEq2}) is the combination of
a homogeneous DC field $\mathbf{E}_{00}$ and the screened electric
field of the SAW, $\mathbf{E}(t,x)=\sum_{\omega q}E_{\omega
q}e^{iqx-i\omega t}\hat{\mathbf{x}}$, found from the unscreened
SAW field via $E_{\omega q}=E_{\omega q}^{\mathrm{SAW}}/\kappa
(\omega, q )$, where $\kappa (\omega, q )$ is the dielectric
function of the whole 2D structure.
The density modulation $n_{\omega q}=\gamma g_{\omega q}^{0}$ 
induced by the SAW sets up an induced field
$E_{\omega q}^{\mathrm{ind}}=-\frac{i2\pi eq}{\chi
|q|}\gamma g_{\omega q}^{0},$ that we take into account
at the level of Thomas-Fermi screening,  
so that $E_{\omega q}=E_{\omega
q}^{\mathrm{SAW}}+E_{\omega q}^{\mathrm{ind}}$. In the analysis of
screening, the DC part of the electric field can be ignored
($f_{00}=0$), and self-consistency yields

\begin{eqnarray}
g_{\omega q}^{0} &=&\frac{eE_{\omega q}}{iq}\frac{1-(1-i\omega
\tau )K}{1-K},
\label{g-naught} 
\end{eqnarray}
and

\begin{equation}
\kappa(\omega, q) =1+\frac{1}{a_{\mathrm{scr}}|q|}\frac{1-\left( 1-i\omega
\tau \right)K}{1-K}.  \label{kappa}
\end{equation}
In the limit that
$qR_c \gg 1$, which is our regime of interest, the dielectric
function becomes

\begin{equation}
\kappa(\omega,q) = 1 + \frac{1-\left( 1-i\omega \tau
\right)K}{a_{\mathrm{scr}}|q|} , \label{kappa2}
\end{equation}
and for most SAWs, $a_{\mathrm{scr}}|q| \ll 1$, 
which implies $\kappa \simeq 1/(a_{\rm scr}|q|)$.
When we account for dispersion, we find that the system
has resonances at 
$\omega=N\omega_c + \Delta_N$ where $\Delta_N$ is the dispersive
shift of the $N^{th}$ resonance. 

We use Eq. (\ref{kappa}) to
find the eigenmodes of the system. Setting $\kappa=0$ and
inserting the expression for $K$, one finds to leading
order in $\omega_c\tau$,

\begin{equation}
\Delta _{N}=\dfrac{N\omega _{c}J_{N}^{2}(qR_{c})}{%
a_{\mathrm{scr}}|q|+ 1-J_{N}^{2}(qR_{c})-NA_{N} }  ,
\label{disper-shift}
\end{equation}
where
\begin{equation}
A_N=\sum_{p=1}^\infty \frac{J^2_{N-p}(qR_c) - J^2_{N+p}(qR_c)}{p}.
\end{equation}
In the limit $qR_c \to 0$ we recover the magneto-plasmon
dispersion, $\Delta _{1}\approx |q|R_{c}^{2}\omega
_{c}/a_{\mathrm{scr}}$, whereas in the limit of $qR_c \gg
1$ screening becomes important and $\Delta_N \ll N\omega_c$. 
The crossover between the
two regimes, which may be evaluated by considering the denominator of
Eq.~(\ref{disper-shift}), occurs at a wavenumber $q_{0}R_c\sim
a_{\mathrm{scr}}/R_{c} \ll 1$. In our discussion below
we take into account the screening of the SAW field. 

\subsubsection{Magnetoresistance oscillations}
\label{sec:sr_mro}
To find the steady state current, we analyze the time/space
average of the kinetic equation in Eq.~(\ref{KinEq}) and take into
account the dynamical perturbation $f_{\omega q}$

\begin{eqnarray}
&{\displaystyle \left[\partial _{\varphi }+\frac{1}{\omega _{c}\tau
}\right]f_{00}-\dfrac{\tau^{-1}-\tau _{in}^{-1}}{\omega _{c}}f_{00}^{0}+\frac{e\mathbf{v\cdot E}_{00}}{%
\omega _{c}}\partial _{\epsilon }f_{T} } \nonumber \\
&\hspace*{0.6cm} 
{\displaystyle = -\sum_{\omega q}\frac{eE_{-\omega -q}}{\omega
_{c}}\left[\mathrm{v}\cos \varphi\, \partial _{\epsilon }-\frac{\sin
\varphi }{p}\partial _{\varphi}\right]f_{\omega q}. } \label{zeroharm}
\end{eqnarray}

In our analysis of classical magnetoresistance
oscillations we assume that $\tau_{in} \gg \tau$ and since
we are not interested in the heating associated with this term we
drop it here.  However it is important for our analysis of the first
quantum correction to the classical result that we present in Sec.~\ref{sec:quantumKE}.
We substitute the solution Eq.~(\ref{GF1}) into
Eq.~(\ref{zeroharm}), keeping track of the effect of the
perturbation of the time/space averaged function $f_{00}$ on
$f_{\omega q}$. This procedure automatically includes
SAW-induced non-linear effects. We
multiply Eq.~(\ref{zeroharm}) by $(2\omega _{c}/e\mathrm{v}_F^2) {\mathrm v}
e^{-i\varphi }$, integrate with respect to $\epsilon $ and
$\varphi $, then use the relation between the $x$ and $y$
components of the DC current, $ j_{x}-ij_{y}=e\gamma
\mathrm{v}_{F}g_{00}^{1}$ and the harmonic $g_{00}^{1}$ (note that
electrical neutrality requires $g_{00}^{0}=0$), which gives

\begin{eqnarray}
&  &{\displaystyle \frac{2\omega _{c}}{e\mathrm{v}_{F}^{2}}\int \frac{d\varphi }{2\pi
}e^{-i\varphi }\int d\epsilon \left\{ \left[ \partial _{\varphi
}+\frac{1}{\omega _{c}\tau }\right]\mathrm{v}f_{00}\right. } \nonumber   \\
& &{\displaystyle \hspace*{0.5cm}
\left. + \, \sum_{\omega q}\frac{\mathrm{v}eE_{-\omega -q}}{\omega
_{c}}\left[\mathrm{ v}\cos \varphi\, \partial _{\epsilon }-\frac{\sin
\varphi }{p}\partial_{\varphi }\right]f_{\omega q}\right\} } \nonumber \\
& = & {\displaystyle
\frac{j_{x}}{\mathrm{v}_{F}^{2}\tau e^{2}\gamma /2} \frac{1}{2}%
\sum_{\omega q}\left\vert \frac{leE_{\omega q}}{\epsilon
_{F}}\right\vert^{2}\frac{K}{1-K}  } \nonumber \\
& &{\displaystyle
\hspace*{1cm}+\left( \frac{i\omega _{c}\tau +1}{\mathrm{v}_{F}^{2}\tau e^{2}\gamma /2}%
\right) [j_{x}-ij_{y}] } \nonumber \\
&
= & E_{00}^{x}-iE_{00}^{y},
\end{eqnarray}
and this can be used to determine the classical SAW induced change of the
resistivity tensor, $\delta ^{c}\hat{\rho}$. The relation between
the electric field and current is
$\mathbf{E}=\hat{\rho}\,\mathbf{j}+\delta ^{c}\hat{\rho}\,\mathbf{
j}$, where $$\hat{\rho} = \rho_0 \left(\begin{array}{cc} 1 & \wct \\ -\wct & 1 \end{array}\right),$$ 
is the Drude resistivity tensor.  Thus we
find the resistivity corrections

\begin{eqnarray}
\frac{\delta ^{c}\rho _{xx}}{\rho _0}
&=&\frac{1}{2}\sum_{\omega q}\left\vert \frac{elE_{\omega
q}}{\epsilon _{F}}\right\vert ^{2}\mathrm{Re}\left\{ \frac{K}{1-K}\right\} ,  \label{WithK} \\
\delta ^{c}\rho _{yy} &=&0,  \notag \\
\dfrac{\delta ^{c}\rho _{yx}}{\rho _0} &=&\dfrac{\delta ^{c}\rho _{xy}}{
\rho _{0}}=-\frac{1}{2}\sum_{\omega q}\left\vert \frac{leE_{\omega q}}{\epsilon _{F}}%
\right\vert ^{2}\mathrm{Im}\left\{\frac{K}{1-K}\right\} =0,
\notag
\end{eqnarray}
and with the use of $E_{\omega q}=E_{\omega
q}^{\mathrm{SAW}}/\kappa (\omega, q) \simeq qa_{scr} E^{\rm SAW}_{\omega q}$, Eq.~(\ref{K-1}), 
$\omega_c\tau \gg 1$  and $qR_c  \gg 1$ we formally justify the result in Eq.~(\ref{Fullresult2}).
The magnetoresistance correction is:

\begin{equation}
\frac{\delta ^{c}\rho _{xx}}{\rho _0} = 2\left( ql\mathcal{E}\right)
^{2}\sum_{N=-\infty }^{\infty }\frac{J_{N}^{2}(qR_{c})}{1+\left( \omega
-N\omega _{c}\right) ^{2}\tau ^{2}} .
\end{equation}

\subsubsection*{Strong damping ($\omega_c\tau \ll 1$)}
At low magnetic fields, there is experimental \cite{Beton} and theoretical \cite{MirlinW,Boggild} 
evidence that there is exponential damping of Weiss oscillations, and it seems natural
that similar behaviour should be observed for SAW-induced oscillations.  We explore this
question and find the functional form of the damping for isotropic scattering in this section,
and for small angle scattering in Sec.~\ref{sec:strongd-sma}.
In the strong damping limit ($\wct \ll 1$), we investigate Eq.~(\ref{K-1}) when $qR_c \gg 1$
and find the values of $\varphi$
and $\tilde{\varphi}$ that lead to a stationary phase. We then integrate over
fluctuations about each point of stationary phase.  These saddle points are $\varphi = \pi/2$
and $\varphi = 3\pi/2$, and $\tilde{\varphi}$ takes values which are any positive or
negative odd integer multiplying $\pi/2$ such that $\tilde{\varphi} \leq \varphi$. 
When we sum the results of integrating about each saddle point,
we get (to lowest order in $e^{-\frac{\pi}{\omega_c\tau}}$)

\begin{eqnarray}
K = \frac{1}{\omega_c\tau} \left[\frac{1}{|qR_c|} + \frac{2}{qR_c}
e^{-\frac{\pi}{\omega_c\tau} + \frac{i\pi\omega}{\omega_c}}\sin(2qR_c)\right].
\label{eq:sd-isotropic}
\end{eqnarray}
In the limit that the magnetic field goes to zero, with $ql \gg 1$, this leads to a 
resistance change

\begin{equation}
\left.\frac{\delta^c\rho_{xx}}{\rho_0}\right|_{B=0} = 2ql {\mathcal E}^2 ,
\end{equation}
and if we consider the resistance change $\delta\rho_{xx}$, we find it is

\begin{equation}
\delta^c\rho_{xx} = \frac{h^2}{\pi e^2} \frac{q}{p_F} {\mathcal E}^2,
\label{eq:noBfield}
\end{equation}
which is independent of disorder. Summation over $\pm \omega$ and taking the imaginary part of
$K$ as in Eq.~(\ref{WithK}) leads to the same result that $\delta^c \rho_{xy} = 0$.

  We interpret this as a SAW-induced backscattering
contribution to the resistance, which will dominate in the limit that the SAW wavelength
is much less than the mean free path.  We ignore the contribution to the resistance from 
channeled orbits, which can also lead to a positive contribution to the magnetoresistance
in the small magnetic field limit. \cite{Menne,Beton2} 
The condition for the existence of these orbits is that the force from the screened SAW field  is 
larger than the Lorentz force, i.e.
${\mathcal E} > \frac{\omega_c}{\omega}\frac{s}{\mathrm{v}_F}$,\cite{MirlinW,Menne} and we assume that
${\mathcal E}$ is sufficiently small for their contribution to be ignored (this is generally the
case over most of the magnetic field range that we show in our figures).

\subsection{Small angle scattering}
\label{sec:long-range-potential}
A long range disorder potential leads to a non-isotropic scattering probability, and this 
implies that there are two scattering times that we need to take into account.  One is
the total scattering time, $\tau_s$,\cite{MirlinW} and the other is the momentum
scattering time, $\tau$, and in GaAs heterostructures, $\tau \gg \tau_s$.  
In the limit that $(qR_c)^2 \ll \frac{\tau}{\tau_s}$, the disorder potential  
leads to diffusion in angle, and can
be studied by replacing the collision integral in the kinetic
equation by a term involving two $\varphi$ derivatives. \cite{MirlinW} 
 The dynamical perturbation of the distribution function, 
Eq.~(\ref{KinHarm}) is thus modified to read

\begin{eqnarray}
\label{eq:kinharm-long1}
\left[ \partial_\varphi - i\frac{\omega}{\omega_c} + iqR_c\cos\varphi - 
\frac{1}{(\wct)^2}\partial_\varphi^2 \right] f_{\omega q} = \tilde{\Psi}(\varphi),
\end{eqnarray}
where
\begin{eqnarray}
\tilde{\Psi}(\varphi)
= - \frac{eE}{\omega_c} \left[ {\mathrm v}\cos\varphi\, \partial_\epsilon -
\frac{\sin\varphi}{p}\partial_\varphi \right] \left(f_{00} +
f_T\right). \label{eq:kinharm-long2}
\end{eqnarray}
In Eq.~(\ref{eq:kinharm-long1}), small angle scattering is introduced in the form
of diffusion along the Fermi surface and is taken into account by the
term $\frac{1}{\tau}\partial_\phi^2 f$.
Now, if we let $f_{\omega q} = h_{\omega q}(\varphi) e^{-iqR_c\sin\varphi}$,
and substitute this into Eq.~(\ref{eq:kinharm-long1}), then solve in the limits
that $qR_c/\wct \ll 1$ 
 and $\omega/\omega_c \ll qR_c$ (the second condition allows
us to ignore the term $\partial_\varphi^2 h$), we can solve the kinetic equation
as before to get 
\begin{eqnarray}
f_{\omega q}(\varphi) = \int^\varphi_{-\infty} d\tilde{\varphi} \,
\tilde{G}(\varphi,\tilde{\varphi}) \tilde{\Psi}(\tilde{\varphi}),
\label{eq:gtilde}
\end{eqnarray}
where
\begin{eqnarray}
\tilde{G}\left(\varphi,\tilde{\varphi} \right) & = &
\exp\bigg\{ iqR_c \left(\sin\tilde{\varphi} - \sin\varphi\right)  \nonumber \\
& &  \hspace*{0.5cm} \left.
+ \left(\frac{1}{\omega_{c}\tau^*} - 
i \frac{\omega}{\omega_c}\right)\left( \tilde{\varphi} - \varphi \right)
\right. \nonumber 
\\ 
& &  \hspace*{0.5cm} \left.-\frac{1}{2\omega_c\tau^*}
\left( \sin 2\varphi -\sin 2\tilde{\varphi}
\right)\right\} , 
\label{eq:A}
\end{eqnarray}
and $\tau^*$ was defined in Eq.~(\ref{eq:taustar}).
Our results for isotropic scattering are
modified by replacing $G$ by $\tilde{G}$. The long-range
potential problem is then reduced to a calculation of $\tilde{K}$ for the
modified Green function, which we define in analogy with Eq.~(\ref{K-1}) as

\begin{eqnarray}
\tilde{K} = 
\int_0^{2\pi} \frac{d\varphi}{2\pi} \int^\varphi_{-\infty} \frac{d\tilde{\varphi}}{\omega_c\tau^*} 
\, \tilde{G}(\varphi,\tilde{\varphi}).
\label{eq:Ktildedef}
\end{eqnarray}
We can write the following exact expression for $\tilde{K}$:
\begin{eqnarray}
\tilde{K} & = & 
\sum_{s,m,p= -\infty}^\infty 
\frac{J_{s+2m-2p}(qR_c)J_s(qR_c)}{ 1 + i((s + 2m)\omega_c - \omega)\tau^*} \nonumber \\
& &  \hspace*{0.3cm} \times \, 
e^{\frac{i\pi}{2}(m+p)}I_m\left(\frac{1}{2\omega_c\tau^*}\right) I_p
\left(\frac{1}{2\omega_c\tau^*}\right),
\label{eq:lrfull}
\end{eqnarray}
where $I_m(x)$ is the modified Bessel function of the first kind.  We study $\tilde{K}$ in the weak
damping ($\omega_c\tau^* \gg 1$) and the strong damping ($\omega_c\tau^* \ll 1$) limits.    

\subsubsection*{Weak damping ($\omega_c\tau^* \gg 1$)}
In the weak damping limit, we can make use of the asymptotic expansion of the modified Bessel 
functions for small argument, and need only retain the $m = p = 0$ terms. 
$\tilde{K}$ has the same form as $K$, except that $\tau$ 
 is replaced by $\tau^*$, i.e. 
\begin{eqnarray}
\tilde{K} = \sum_{N=-\infty }^{\infty }\dfrac{J_{N}^{2}(qR_{c})}{1+i
(N\omega_{c}-\omega )\tau^*}\quad  ,
\label{eq:Ktilde} 
\end{eqnarray}
where $\frac{1}{\tau^*} = \frac{(qR_c)^2}{2\tau}$ was introduced in Eq.~(\ref{eq:taustar}).

\subsubsection*{Strong damping ($\omega_c\tau^* \ll 1$)}
\label{sec:strongd-sma}
In the strong damping limit when $qR_c \gg 1$ and $(qR_c)^2 \ll \frac{\tau}{\tau_s}$, we investigate Eq.~(\ref{eq:Ktildedef})
and use a similar saddle point procedure to the one we used for strong damping in the case of
isotropic scattering.  After adding the contributions from
integrating about each saddle point,
we get (to lowest order in $e^{-\frac{\pi}{\omega_c\tau^*}}$)

\begin{eqnarray}
\tilde{K} = \frac{1}{\omega_c\tau^*} \left[\frac{1}{|qR_c|} + \frac{2}{qR_c}
e^{-\frac{\pi}{\omega_c\tau^*} + \frac{i\pi\omega}{\omega_c}}\sin(2qR_c)\right].
\end{eqnarray}
Note that this is the same form as the expression 
for $K$ for isotropic scattering in the limit
$\omega_c\tau \ll 1$, except that $\tau$ replaces $\tau^*$.
If we want to investigate the limit in which the magnetic field goes to zero,
then the kinetic equation [Eq.~(\ref{eq:kinharm-long1})] as written previously
is inapplicable when $(qR_c)^2 \gg \frac{\tau}{\tau_s}$.\cite{MirlinW}  In our calculation above, there is a
damping factor of $D = e^{-\frac{\pi}{\omega_c\tau^*}}$, which arises naturally
as a result of our saddle point analysis.
In Ref.~\onlinecite{MirlinW} an alternative approach was used to calculate the 
damping factor in the low magnetic field limit.  In that approach the damping factor above 
is the high field limit of

\begin{equation}
D_1 = \exp\left\{-\frac{\pi}{\omega_c\tau_s}\left[1 - 
\frac{1}{\sqrt{1 + \frac{\tau_s}{\tau}(qR_c)^2}}
\right] \right\},
\label{eq:mirlin}
\end{equation}
where $\tau_s$ is the total relaxation rate.  If, as in Sec.~\ref{sec:qual-sma}, we assume
that phase and guiding center corrections are uncorrelated, we can calculate a second damping
factor associated with phase corrections in addition to the guiding center contribution in
Eq.~(\ref{eq:mirlin}), using the same method.  This gives

\begin{equation}
D_2 = \exp\left\{-\frac{\pi}{\wct_s} \frac{\left(\frac{\omega}{\omega_c}\right)^2}{1+
\left(\frac{\omega}{\omega_c}\right)^2 \frac{\tau_s}{\tau}}\right\},
\end{equation}
and at moderate fields, we may approximate $D = D_1 D_2$, so that 

\begin{eqnarray}
\tilde{K} = \frac{1}{\omega_c\tau^*} \left[\frac{1}{|qR_c|} + \frac{2}{qR_c}
D e^{\frac{i\pi\omega}{\omega_c}}\sin(2qR_c)\right].
\end{eqnarray}

If we assume that the two effects are correlated (as we should in the $(qR_c)^2 \gg
\frac{\tau}{\tau_s}$ limit), 
then we get the expression

\begin{equation}
D = \exp\left\{ -\frac{1}{\wct_s} \int_0^{\pi} d\phi \frac{(ql\sin\phi - \omega\tau)^2}{
(ql\sin\phi - \omega\tau)^2 + (\wct) (\wct_s)}\right\},
\end{equation}
which implies that in the $\omega_c\tau \to 0$ limit, with $ql \gg \omega\tau \gg 1$ and 
$\frac{\tau}{\tau_s} \gg 1$

\begin{eqnarray}
D \simeq \exp\left\{ - \frac{\pi}{\wct_s} \left[ 1 - \frac{\omega\tau}{(ql)^2} \wct 
\sqrt{\frac{\tau_s}{\tau}}\right]\right\},
\end{eqnarray}
which gives $\lim_{\omega_c \to 0} D = e^{-\frac{\pi}{\wct_s}} \neq
\lim_{\omega_c \to 0} D_1 D_2 = e^{-\frac{2\pi}{\wct_s}}$.

\subsubsection{Screening}

In the above discussion, 
the solutions obtained for $f_{\omega q}$ in Sec.~\ref{sec:short-range-potential} 
and for $f_{\omega q}$ when there is  small-angle
scattering differ in that the kinetic equation for isotropic
scattering contained the term $g^0_{\omega q}/\omega_c\tau$, which
is absent here. When we solve 
for $g^0_{\omega q}$, and hence the 
dielectric function, Eq.~(\ref{kappa}), in the presence of small-angle 
scattering, we get

\begin{eqnarray}
g_{\omega q}^{0} &=&\frac{eE_{\omega q}}{iq}\left[1-(1-i\omega
\tau^* )\tilde{K}\right],
\label{eq:g-naught2}
\end{eqnarray} 
which implies a dielectric function
\begin{equation}
\kappa(\omega,q) =1+\frac{1-\left( 1-i\omega \tau^*
\right)\tilde{K}}{a_{\mathrm{scr}}|q|} ,
\label{eq:kappa3}
\end{equation}
which is very similar to Eq.~(\ref{kappa2}) with $\tilde{K}$
replacing $K$.  When $qR_c \gg 1$, screening of SAWs is identical
for both short and long range potentials.  Analysis of
$\Delta_N$ and $A_N$ as in Sec.~\ref{sec:screening} leads to
the same $q_0$ at which screening becomes important.

\subsubsection{Magnetoresistance oscillations}
In this section we derive the magnetoresistance
oscillations analogously to Sec.
\ref{sec:sr_mro}. The system of equations that we
wish to solve is

\begin{eqnarray}
&{\displaystyle \partial_\varphi f_0 + \frac{eE_{\omega q}}{\omega_c} 
\left[ {\mathrm v}\cos\varphi \, \partial_\epsilon - \frac{\sin\varphi}{p} \partial_\varphi 
\right] f_{\omega q}  }\nonumber \\
& {\displaystyle \hspace*{0.5cm} =  -\frac{e\bvec{v}\cdot\bvec{E}_0}{\omega_c} \partial_\epsilon f_T + 
\frac{1}{\omega_c\tau}\partial^2_\varphi f_0 } ,
\label{eq:sma1}
\end{eqnarray}
in combination with Eqs.~(\ref{eq:kinharm-long1}) and (\ref{eq:kinharm-long2}).
The solution of Eq. (\ref{eq:kinharm-long1}) for $f_{\omega q}$, is
shown in Eq.~(\ref{eq:gtilde}). We take the equation for $f_0$, and
integrate with respect to energy and $\varphi$ after multiplying by ${\mathrm v}e^{-i\varphi}$,
as before, and obtain

\begin{eqnarray}
\delta E^x_{00} - i\delta E^y_{00} & = & \frac{2\omega_c}{e m^2 {\mathrm v}_F^3} 
\left(\frac{eE_{\omega q}}{\omega_c}\right)^2  (g^1_0 + g^{-1}_0) \nonumber 
 \\ & & \hspace*{0.3cm} \times
\int_0^{2\pi} \frac{d\varphi}{2\pi} \int^\varphi_{-\infty} d\tilde{\varphi} \, 
\tilde{G}(\varphi,\tilde{\varphi}) , \nonumber 
\end{eqnarray}
which is the same as we found for isotropic scattering, except 
that $\tilde{G}$ replaces $G$. The resistivity correction is thus

\begin{eqnarray}
\frac{\delta^c\rho_{xx}}{\rho_0} = \frac{1}{2}\sum_{\omega q} \left|\frac{elE_{\omega q}}{\epsilon_F}
\right|^2 \frac{\tau^*}{\tau} {\rm Re}\left\{\tilde{K}\right\} .
\label{eq:fullrholr}
\end{eqnarray}
As in the case of short-range scattering there are no corrections to any other 
components of the resistivity tensor.  Unlike the case of isotropic scattering
there is {\it not} a factor of $1 - \tilde{K}$ in the denominator of 
Eq.~(\ref{eq:fullrholr}) -- this is because the resistivity correction
is linear in $g^0_{\omega q}$, which is modified for small-angle scattering [see
Eq.~(\ref{eq:g-naught2})].

Thus our results for the classical contribution to the magnetoresistance can be 
summarized as follows.
For weak damping, $\omega_c\tau^* \gg 1$, and $1 \ll (qR_c)^2 \ll \frac{\tau}{\tau_s}$ 
the resistivity correction is:

\begin{equation}
\frac{\delta^c \rho _{xx}}{\rho_0}=\frac{2{\mathrm v}_{F}^{2}\omega^{2}\tau \tau^*}{
s^{2}}{\mathcal{E}}^{2}\sum_{N=-\infty }^{\infty }\frac{J_{N}^{2}(qR_{c})}{%
1+\left( \omega -N\omega _{c}\right) ^{2}{\tau^*}^2}  \, .
 \label{eq:Fullresult-longr}
\end{equation} 
For strong damping, $\omega_c\tau^* \ll 1$, and $1 \ll (qR_c)^2 \ll \frac{\tau}{\tau_s}$,

\begin{eqnarray}
\frac{\delta^c \rho_{xx}}{\rho_0} = 2 ql {\mathcal E}^2 
\left[ 1 + 2e^{-\frac{\pi}{\omega_c \tau^*}} \cos\left(\frac{\pi\omega}{\omega_c}\right) 
\sin(2qR_c)\right], \nonumber \\
\label{eq:strongd}
\end{eqnarray}
whilst for $\omega_c\tau^* \ll 1$ and $(qR_c)^2 \gg \frac{\tau}{\tau_s}$,

\begin{eqnarray}
\frac{\delta^c \rho_{xx}}{\rho_0} = 2 ql {\mathcal E}^2 
\left[ 1 + 2e^{-\frac{\pi}{\omega_c \tau_s}} \cos\left(\frac{\pi\omega}{\omega_c}\right) 
\sin(2qR_c)\right]. \nonumber \\
\label{eq:strongd2}
\end{eqnarray}
Summation over $\pm \omega$ and $\pm q$ for the imaginary part of $\tilde{K}$ ensure
that $\delta^c\rho_{xy}$ and $\delta^c\rho_{yy}$ vanish as for short range scattering [see Eq.~(\ref{WithK})].
The three resistivity regimes identified above correspond respectively to high 
[Eq.~(\ref{eq:Fullresult-longr})], intermediate [Eq.~(\ref{eq:strongd})], and
low [Eq.~(\ref{eq:strongd2})] magnetic fields.  The crossover between high and intermediate magnetic fields is at
$\omega_c\tau^* \sim 1$, which is equivalent to $\omega_c\tau \sim (ql)^\frac{2}{3}$,
and the crossover between intermediate and low magnetic fields is when $(qR_c)^2 \sim \frac{\tau}{\tau_s}$,
which is equivalent to $\omega_c\tau \sim ql\sqrt{\frac{\tau_s}{\tau}}$. 
The number of damped oscillations at low magnetic field has recently been calculated elsewhere.\cite{Robinson2}
As for short range scattering, these results are for the regime where channeled orbits can be ignored, 
which requires that the Lorentz force should be stronger than the screened SAW field, i.e. ${\mathcal E}
 < \frac{\omega_c}{\omega}\frac{s}{\mathrm{v}_F}.$

Both Eqs.~(\ref{eq:Fullresult-longr}) and (\ref{eq:strongd}) 
reduce to the results 
found by Mirlin and W\"{o}lfle \cite{MirlinW} in the limit that $\omega \to 0$, provided
one replaces $\omega/s$ by $q$ in Eq.~(\ref{eq:Fullresult-longr}).
To extrapolate this result to lower magnetic fields, the best we can do is a similar
procedure to Mirlin and W\"{o}lfle, which is to replace $e^{-\frac{\pi}{\omega_c \tau^*}}$
by the damping factor $D$ discussed in Sec.~\ref{sec:strongd-sma}, 
leading to Eq.~(\ref{eq:strongd2}).  In the limit that 
$B \to 0$ this reduces to the same behaviour as in the case of isotropic scattering, and
the magnetoresistance correction is given by Eq.~(\ref{eq:noBfield}).  We neglect the possibility
of channeled orbits, as discussed in Sec.~\ref{sec:sr_mro}.

\begin{figure}[h]
\center{\includegraphics[width=6cm,angle=270]{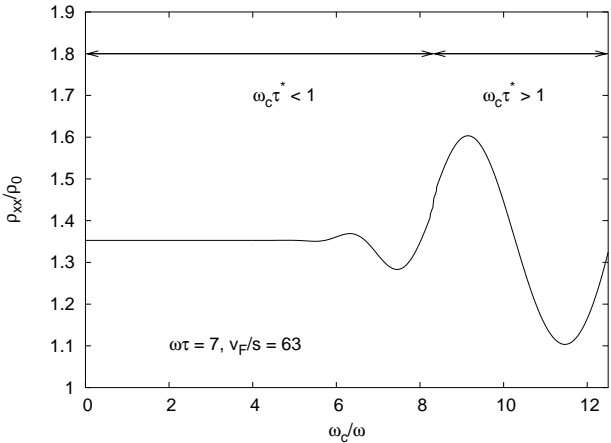}}
\caption{Magnetoresistance correction for small-angle scattering, with 
$\omega\tau=7$, ${\mathrm v}_F/s= 63$, and ${\mathcal E} = 0.02$.  The curve is determined 
by splicing the expressions for
the resistance in the strong and weak damping limits ($\wct^* < 1$ and $\wct^* > 1$
respectively), which are indicated in the figure.}
\label{fig:smalla0}
\end{figure}

\begin{figure}[h]
\center{\includegraphics[width=6cm,angle=270]{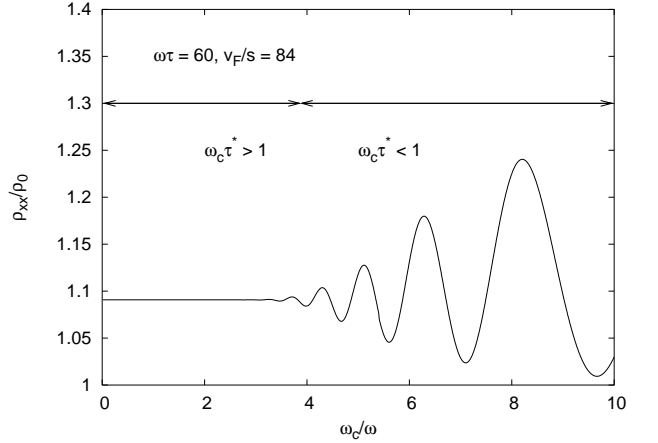}}
\caption{Magnetoresistance correction for small-angle scattering, with 
$\omega\tau=60$, ${\mathrm v}_F/s = 84$, and ${\mathcal E} = 0.003$.  The curve is determined 
by splicing the expressions for
the resistance in the weak and strong damping limits, which are indicated in the figure. }
\label{fig:dualsa}
\end{figure}

\begin{figure}[h]
\center{\includegraphics[width=6cm,angle=270]{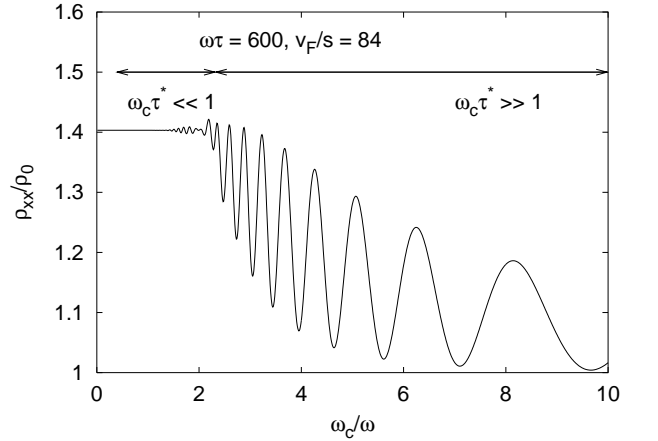}}
\caption{Magnetoresistance correction for small-angle scattering, with 
$\omega\tau=600$, ${\mathrm v}_F/s = 84$, and ${\mathcal E} = 0.002$. 
The curve is a splice of the formulae
for the strong and weak damping limits, and the range of validity for each region is indicated. }
\label{fig:smalla2}
\end{figure}

In Figs.~\ref{fig:smalla0}, \ref{fig:dualsa}, and \ref{fig:smalla2} we show the behaviour
of the magnetoresistance as determined by
splicing the results of Eq.~(\ref{eq:Fullresult-longr}) and Eq.~(\ref{eq:strongd})
for the weak and strong damping cases respectively.  In  Fig~\ref{fig:smalla0} we use the same 
sample parameters as in Fig.~\ref{fig:zudov1}, whilst both Figs.~\ref{fig:dualsa} and \ref{fig:smalla2}
are for ${\mathrm v}_F/s = 84$, with 
$\omega\tau = 60$ in Fig.~\ref{fig:dualsa} and $\omega\tau = 600$ in Fig.~\ref{fig:smalla2};
for the sample parameters in Ref.~\onlinecite{Zudov2} these correspond to $\omega = 2\pi\times 10$ GHz and
$2\pi\times 100$ GHz  respectively. 
Note that, owing to our requirement $qR_c\ll \omega_c\tau$, 
our results should not be trusted in the region $\omega_c/\omega 
\lesssim \sqrt{{\mathrm v}_F/(s \omega\tau)}$ (which is of order 1 for the parameters used), i.e. 
at low magnetic fields.
However, our expectation is that magnetoresistance oscillations should be strongly
damped in this regime, as described in Eq.~(\ref{eq:strongd2}).
We also note that for $\omega_c/\omega \gg 1$, the resistance correction 
in the case of small-angle scattering is enhanced over that expected for isotropic
scattering with the same value of the transport time (i.e. the same 2DEG mobility).

\section{Quantum Kinetic contribution to SAW-induced magnetoresistance}
\label{sec:quantumKE}

The absorption of SAWs by electrons in two dimensions changes their steady state
distribution over energy. Characteristics that are independent of energy cause
no additional changes to magnetoresistance beyond those previously
described.   However, when  $\omega _{c}\tau
\gg 1$, Landau level quantization is present, leading to an
oscillatory energy-dependence of the electron DOS for overlapping
Landau levels of $\tilde{\gamma}(\epsilon
)=\left[ 1-\Gamma \cos \left( 2\pi \epsilon /\hbar \omega _{c}\right) \right]
\gamma $, where $\Gamma = 2e^{-\frac{\pi}{\omega_c\tau_q}} \ll 1$, and $\tau_q$ 
is the quantum lifetime of the Landau levels. \cite{Ando}  [A calculation of
the density of states for a long-range disorder potential\cite{Raikh}
shows that $\tau_q$ can be field-dependent.]
This gives rise to a 
quantum contribution to the geometrical
commensurability oscillations which can persist up to high temperatures. The
oscillations in the DOS impose 
oscillations on the electron elastic scattering rate, $%
\tau ^{-1}(\epsilon )=\tau ^{-1}\tilde{\gamma}(\epsilon)/\gamma $, which in turn gives 
a contribution to the observable conductivity.
At low temperatures, $k_{B}T\lesssim \hbar \omega _{c}$, the DOS
oscillations lead to Shubnikov-de Haas (SdH) oscillations in conductivity. At high
temperatures $k_{B}T\gg \hbar \omega _{c}$, thermal broadening smears out the SdH
oscillations, but the quantum contribution can remain as a non-linear effect
after energy averaging.

We follow a similar approach to Dmitriev {\it et al.} \cite{Mirlin,Aleiner}
to study the quantum kinetic equation to obtain the first 
quantum correction to the classical magnetoresistance induced by SAWs.    
This change in the distribution is oscillatory in energy and leads to a contribution to 
the DC conductivity that oscillates as a function of $\omega/\omega_c$.  The effect
depends on the efficiency of energy relaxation and hence
is proportional to $\tau_{in}$; this effect 
dominates the quantum corrections discussed in Refs.~\onlinecite{Durst} when $\tau_{in}/\tau_q \gg 1$.
If this is the case, the energy relaxation  time is long compared to the Landau level lifetime, 
allowing a strongly non-equilibrium electron distribution as a function of
energy to arise.

The photoconductivity $\sigma_{\rm ph}$ determines the longitudinal current flowing
in response to a DC electric field in the presence of SAWs:

\begin{equation}
\bvec{j} \cdot \bvec{E}_{00} = \sigma_{\rm ph} |\bvec{E}_{00}|^2,
\end{equation}
and can be related to the photoresistivity by $\rho_{\rm ph} \simeq \rho_{xy}^2\sigma_{\rm ph}$,
where $\rho_{xy} = eB/n_e$.  To calculate the photoconductivity, one integrates over the
distribution function 

\begin{equation}
\sigma_{\rm ph} = 2\int d\epsilon \, \sigma_{00}(\epsilon)\left[-\partial_\epsilon f(\epsilon)\right],
\label{eq:photoc}
\end{equation}
and in the leading approximation,

\begin{equation}
\sigma_{00}(\epsilon) = \sigma_{00}^D\frac{\tilde{\gamma}^2(\epsilon)}{\gamma^2},
\end{equation}
where 

\begin{equation}
\sigma^D_{00} = \frac{e^2\gamma {\mathrm v}_F^2}{2\omega_c^2 \tau}   ,
\label{eq:drudedef}
\end{equation}
is the Drude conductivity.
In obtaining a solution to the problem, we are only interested in effects due
to the non-trivial energy dependence of the electron distribution function $f(\epsilon)$.

To perform this calculation we need the classical dynamical conductivity, from which 
we consider the energy dependence of the density of states $\gamma$ and the momentum
relaxation time $\tau$.  We calculate the classical dynamical conductivity below in 
Sec.~\ref{sec:classdyn}, and then use it to derive the quantum energy balance 
equation and magnetoresistance correction in Sec.~\ref{sec:qenergybalance}.

\subsection{Classical dynamical conductivity}
\label{sec:classdyn}
The magnetic field dependence of the resistivity change reflects the form of
the SAW attenuation by the 2D electrons determined by the real part of the
longitudinal dynamical conductivity $\sigma _{\omega q}$, 
\begin{equation}
\mathrm{Re}\left\{\sigma _{\omega q}\right\}=\gamma s^{2}\tau e^{2}\,\mathrm{Re}\left\{ 
\frac{K}{1-K}\right\} \mathrm{.}  
\label{SigmaOmegaQ}
\end{equation}
We obtain $\sigma_{\omega q}$ from considering attenuation in the form
\begin{equation}
\left\langle \mathbf{E}_{\omega q}\cdot \mathbf{j}_{\omega q}\right\rangle = \sigma_{\omega q} |E_{\omega q}|^2 ,
\nonumber 
\end{equation}
and we use the expression in Eq.~(\ref{SigmaOmegaQ}) in the following subsection.  In 
analogy with our previous work, we find that the equation for $\sigma_{\omega q}$ for
small-angle scattering is

\begin{equation}
\mathrm{Re} \left\{\sigma _{\omega q}\right\}=\gamma s^{2}\tau^* e^{2}\,\mathrm{Re}\left\{ 
\tilde{K}\right\}  .
\label{SigmaOmegaQ2}
\end{equation}

\subsection{Quantum energy balance equation}
\label{sec:qenergybalance}
The quantum energy balance equation in Ref.~\onlinecite{Aleiner} is stated without
detailed derivation. Here we provide a simple derivation as an alternative to the approach used in Ref.~\onlinecite{Aleiner}.
If we consider the energy balance in a 2DEG due to absorption and emission of
SAWs, then the rate at which energy is added is
$ Q(t) =  \bvec{j}_{\omega q}(t)\cdot\bvec{E}_{\omega q}(t),$
and $\bvec{j}_{\omega q}(t) = \sigma_{\omega q} \bvec{E}_{\omega q}(t)$, which when we average over
time gives
$ \bar{Q} =\frac{1}{2}\sigma_{\omega q} |\bvec{E}_{\omega q}|^2 $ 
We can express the rate of absorption or emission 
processes involving energy levels $i$ and $j$ as
\begin{equation}
\Gamma_{ij} = \frac{2\pi}{\hbar}|M_{ij}|^2 \delta(\epsilon_i - \epsilon_j \pm \hbar\omega) ,
\end{equation}
using Fermi's golden rule, where $M_{ij}$ is the matrix element between states $i$ and $j$.  
We assume that $M_{ij} = M$ and calculate the total
power absorption for the classical case ($\gamma$ constant), 
which we equate to $\bar{Q} = \frac{1}{2}\sigma_{\omega q} |\bvec{E}_{\omega q}|^2.$
Hence

\begin{equation}
\bar{Q} = \frac{2\pi}{\hbar} |M|^2  \hbar\omega \gamma (\hbar\omega\gamma),
\end{equation}
where one factor of $\hbar\omega$ is the energy added per
photon absorbed, $\gamma$ is the final density of states, and $ \hbar\omega\gamma$ is the number of electrons below the Fermi level that are available to make a transition.
This gives a formula for $|M|^2$:
\begin{equation}
|M|^2 = \frac{\hbar \bar{Q}}{2\pi (\hbar\omega\gamma)^2}.
\end{equation}

We now consider the change in the number of electrons with energy $\epsilon\to \epsilon+ d\epsilon$
allowing the
 density of states to be energy dependent, $\tilde{\gamma}(\epsilon)$:
\begin{eqnarray}
\begin{array}{l}
{\displaystyle \frac{d}{dt}\left[f(\epsilon)\tilde{\gamma}(\epsilon)d\epsilon\right] } \\
 {\displaystyle =  \frac{2\pi}{\hbar} |M|^2 \left\{ \sum_{\pm} f(\epsilon \pm \hbar\omega) 
\tilde{\gamma}(\epsilon \pm \hbar\omega) d\epsilon \, \tilde{\gamma}(\epsilon)
(1 - f(\epsilon)) \right. } \\
\hspace*{2cm} {\displaystyle - f(\epsilon) \tilde{\gamma}(\epsilon) d\epsilon \,
\tilde{\gamma}(\epsilon \pm \hbar\omega) 
(1 - f(\epsilon \pm \hbar\omega)) \Bigg\} }  \\
\hspace*{0.3cm} 
{\displaystyle - \frac{f(\epsilon) - f_T(\epsilon)}{\tau_{in}}\tilde{\gamma}(\epsilon)d\epsilon. } \end{array} 
 \nonumber \\
\label{eq:qkin1}
\end{eqnarray}
When we require that the distribution be stationary, the left hand side of the equation
vanishes and when we divide by $\tilde{\gamma}(\epsilon)$ we get 

\begin{equation}
\frac{\sigma_{\omega q} |E_{\omega q}|^2}{2\hbar^2\omega^2 \gamma^2} \sum_{\pm}
\tilde{\gamma}(\epsilon \pm \hbar\omega)\left[f(\epsilon \pm \hbar\omega) - f(\epsilon)\right]
= \frac{\left[f(\epsilon) - f_T(\epsilon)\right]}{\tau_{in}}.
\end{equation}
If we want to include the effects of a DC field as well, then we can see the form of the DC term
from the $\omega \to 0$ limit of the right hand side of Eq.~(\ref{eq:qkin1}), 
and we get as a final result the balance equation from Ref. \onlinecite{Aleiner},

\begin{eqnarray}
\frac{\sigma_{\omega q} |E_{\omega q}|^2}{2\hbar^2\omega^2 \gamma^2} \sum_{\pm}
\tilde{\gamma}(\epsilon \pm \hbar\omega)\left[f(\epsilon \pm \hbar\omega) - f(\epsilon)\right]
\nonumber \\
+ \frac{|E_{00}|^2 \sigma_{00}}{\tilde{\gamma}(\epsilon)} \frac{\partial}{\partial\epsilon}
\left[\frac{\tilde{\gamma}(\epsilon)^2}{\gamma^2} \frac{\partial}{\partial\epsilon} 
f(\epsilon)\right] 
= \frac{\left[f(\epsilon) - f_T(\epsilon)\right]}{\tau_{in}}.
\end{eqnarray}
We could also have obtained the same equation by treating $\sigma_{\omega q}$ as energy dependent
and then pulling out factors of the energy dependent 
density of states from $\tilde{\gamma}(\epsilon)$ and $\tau^{-1}(\epsilon + \hbar\omega)$, which is
equivalent to the approach in Ref.~\onlinecite{Mirlin}.  This second
approach works provided $|N - \omega/\omega_c| \leq 1/\wct$ (i.e. we can ignore the
1 in the denominator of the sum over $N$ harmonics in $\sigma_{\omega q}$). At frequencies 
closer to resonance than this, we can no longer ignore the 1 in the denominator and 
the classical expression lacks accuracy.  However, for large $\wct$ this is a relatively small 
region of magnetic fields, and is not relevant at the 
magnetic fields at which ZRS mimima are observed.

We use the expression for $\sigma_{\omega q}$ from Eq.~(\ref{SigmaOmegaQ}) (short-range disorder
potential) or 
Eq.~(\ref{SigmaOmegaQ2}) (long-range disorder potential), which contains the geometric and
frequency commensurability oscillations, and $\sigma_{00}^D$ we know
as the Drude resistivity [Eq.~(\ref{eq:drudedef})].
Introduce the following quantities

\begin{eqnarray}
\label{eq:Pdef}
{\mathcal P} & = & \tau_{in} \frac{2\sigma_{\omega q} |E_{\omega q}|^2}{\hbar^2 \omega^2 \gamma}  , \\
{\mathcal Q} & = & \tau_{in} \frac{4\pi^2 \sigma^D_{00} |E_{00}|^2 }{\hbar^2\omega_c^2 \gamma}  .
\end{eqnarray}
We are then left  with the following equation to solve 

\begin{eqnarray}
&{\displaystyle \frac{{\mathcal P}}{4} \sum_\pm \left(1 - \Gamma 
\cos\left(\frac{2\pi(\epsilon \pm \hbar\omega)}{\hbar\omega_c}\right)\right)
\left(f(\epsilon \pm \hbar\omega) - f(\epsilon)\right) }\nonumber \\
&{\displaystyle + \frac{\hbar^2\omega_c^2 \gamma}{4\pi^2 \tilde{\gamma}(\epsilon)}{\mathcal Q}
\frac{\partial}{\partial\epsilon} \left[
\left(1 - \Gamma \cos\left(\frac{2\pi\epsilon}{\hbar\omega_c}\right)\right)^2
\frac{\partial}{\partial\epsilon} f(\epsilon)\right] } \nonumber \\ 
&= f(\epsilon) - f_T(\epsilon). \nonumber \\
\end{eqnarray}
 
Now, let $f(\epsilon) = f_0(\epsilon) + \Gamma f_1(\epsilon) 
\cos\left[\frac{2\pi \epsilon}{\hbar\omega_c}\right]$,
where $f_0(\epsilon) \simeq f_T(\epsilon)$, so $f(\epsilon) \simeq f_T(\epsilon) + f_{\rm osc}(\epsilon)$.
When we do this, and then sum over $\omega$ and neglect all derivatives of $f_1$ that arise
from Taylor expanding $f(\epsilon\pm \hbar\omega)$ (since this is assumed
to be a smooth function on the energy scale of $k_B T$), and retain only lowest
order terms in $\Gamma$, we get a linear equation in $f_{\rm osc}$ which is trivial to solve and leads
to (noting that $\partial_\epsilon^2 f_T$ 
is also ignored) an identical result to Ref.~\onlinecite{Aleiner}:

\begin{eqnarray}
f_{\rm osc} = \frac{\hbar\omega_c}{2\pi}\frac{\Gamma}{2} \frac{\partial f_T}{\partial \epsilon} 
\sin\left(\frac{2\pi \epsilon}{\hbar\omega_c}\right) \frac{\frac{2\pi\omega}{\omega_c}{\mathcal P}
\sin\left(\frac{2\pi\omega}{\omega_c}\right) + 4{\mathcal Q}}{1 + 
{\mathcal P}\sin^2\left(\frac{\pi\omega}{\omega_c}\right) + {\mathcal Q}}  .
\end{eqnarray}
We then use the expression for the photoconductivity, Eq.~(\ref{eq:photoc})
and get the isotropic change in the DC conductivity

\begin{eqnarray}
\frac{\sigma_{\rm ph}}{\sigma_{00}^D} = 1 + \frac{\Gamma^2}{2} \left[
1 - \frac{\frac{2\pi\omega}{\omega_c}{\mathcal P}\sin\left(\frac{2\pi\omega}{\omega_c}\right) + 4{\mathcal Q}}{1 +
{\mathcal P}\sin^2\left(\frac{\pi\omega}{\omega_c}\right) + {\mathcal Q}}\right] .
\label{eq:sigmaratio}
\end{eqnarray}

Note that in the approach outlined above we made no detailed 
assumptions about the momentum relaxation apart from $\tau_{in} \gg \tau$, such that the
momentum relaxation is efficient in making the distribution isotropic.  Thus  the only 
difference in the expression for a short or long-ranged potential is in the form of
$\sigma_{\omega q}$, which enters in the expression for ${\mathcal P}$ in Eq.~(\ref{eq:Pdef}).
We now present numerical calculations of the resistivity change derived above 
in the presence of SAWs for both isotropic and small-angle
scattering. All these calculations are for the case ${\mathcal Q} = 0$.

\subsection{Numerical results}
In this section we show the resistivity as it is modified by the
quantum correction alone.  We discuss the situation where both classical and 
quantum corrections are important in Sec.~\ref{sec:classquant}.

\subsubsection{Isotropic scattering}
\begin{figure}[h]
\center{\includegraphics[width=6
cm,angle=270]{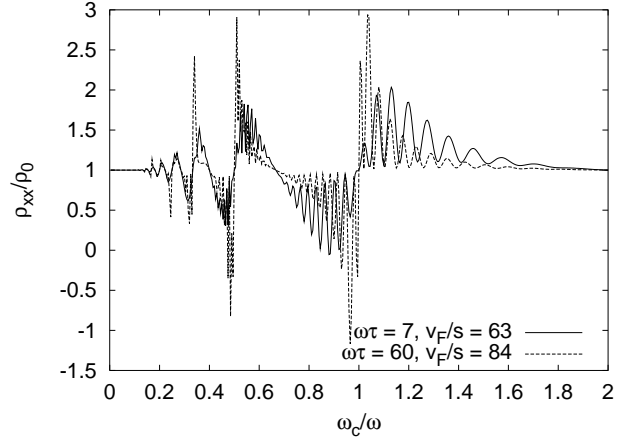}}
\caption{Quantum contribution to the magnetoresistance for $\omega\tau = 7$, ${\mathrm v}_F/s = 63$ and $\omega\tau = 60$, ${\mathrm v}_F/s = 84$, 
both with $\omega\tau_q = 6$, ${\mathcal E} = 0.0015$, and $\omega\tau_{in} = 302$.}
\label{fig:quant}
\end{figure}

We illustrate numerical results for the magnetoresistance oscillations due to the
combination of SAWs and density of states modulations in Fig. \ref{fig:quant}.
We use Eq.~(\ref{K-1}) for $K$ and do not ignore the 1 in the
denominator of the summand.  For  large $\omega\tau$, this should give accurate results except for
a very small range of magnetic fields around each of the resonances where $\omega/\omega_c$
equals an integer.
Note that we divide the expression in Eq.~(\ref{eq:sigmaratio}) by $1 + \Gamma^2 / 2$,
to compare the resistivity with and without SAWs, assuming that there are
density of states modulations in both cases. 
To connect to our previous notation, we note that 

\begin{equation}
{\mathcal P} = 2 \left(\frac{\epsilon_F}{\hbar\omega}\right)^2 (\omega\tau)(\omega\tau_{in}) 
{\mathcal E}^2 {\rm Re}\left\{\frac{K}{1-K}\right\} .
\label{eq:Pconvert}
\end{equation}

In Fig.~\ref{fig:quant} we show the change in the 
magnetoresistance for $\omega\tau_q = 6$ and $\omega\tau_{in} = 302$ 
(using the value of $\tau_{in} = 7.64 \times 10^{-10}$ s 
estimated in Ref.~\onlinecite{Aleiner}) 
for the parameters that we 
calculated the classical magnetoresistance corrections displayed in 
Figs.~\ref{fig:zudov1} and \ref{fig:zudov2}.   As for the case of a microwave field, 
there are peaks in the resistance when $\omega/\omega_c$ is close to an integer, but
with $\delta\rho_{xx} = 0$ when $\omega/\omega_c$ is exactly equal to an integer.
The peaks (and dips) are modulated by geometric commensurability oscillations,
and this leads to a new class of ZRS, of the type predicted for $\omega \ll \omega_c$
in Ref.~\onlinecite{Robinson}. For $\omega\tau = 7$, ${\mathrm v}_F/s = 63$, the ZRS occur for ${\mathcal E} 
\gtrsim 0.0015$ and for $\omega\tau = 60$, ${\mathrm v}_F/s = 84$, ZRS occur for ${\mathcal 
E} \gtrsim 0.001$.

At large $\omega_c/\omega \gg 1$ (not shown) the resistance change is
negative as we predicted analytically previously.\cite{Robinson}
However, the magnitude of the resistance change is very much smaller
than when $\omega/\omega_c \gtrsim 1$, so a very large value of
${\mathcal E}$ would be required to attain ZRS at $\omega_c/\omega \gg
1$.

\subsubsection{Small angle scattering}

\begin{figure}[h]
\center{\includegraphics[width=6
cm,angle=270]{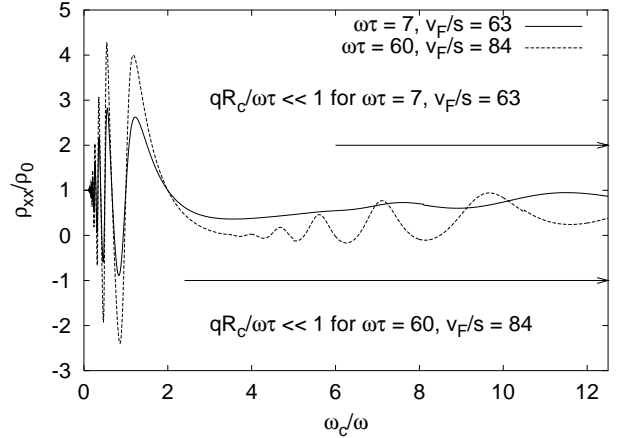}}
\caption{Quantum contribution to magnetoresistance for $\omega\tau = 7$, ${\mathrm v}_F/s = 63$
and for $\omega\tau = 60$, ${\mathrm v}_F/s = 84$.  
In both cases $\omega\tau_q = 6$, ${\mathcal E} = 0.003$, and $\omega\tau_{in} = 302$.
The arrows indicate the range of validity of the results.}
\label{fig:longshortcomp}
\end{figure}

In Fig.~\ref{fig:longshortcomp} we show data for 
the quantum magnetoresistance correction when there is
small-angle scattering for the same sample parameters
and frequencies as in Figs.~\ref{fig:smalla0} and \ref{fig:dualsa}.  Similarly to 
the short-range case, we can relate ${\mathcal P}$ to our previous notation:

\begin{equation}
{\mathcal P} = 2 \left(\frac{\epsilon_F}{\hbar\omega}\right)^2 (\omega\tau^*)(\omega\tau_{in}) 
{\mathcal E}^2 {\rm Re}\left\{\tilde{K}\right\} ,
\end{equation}
where the change from Eq.~(\ref{eq:Pconvert}) reflects that we use a different $\sigma_{\omega q}$
in the two cases.

If $\omega_c/\omega \lesssim 1$ then there are ZRS of the type that are usually
observed with microwaves, however there are another class of magnetoresistance oscillations
at larger values of $\omega_c/\omega$, which are 
of the sort we predicted in Ref.~\onlinecite{Robinson}, and are much less
strongly damped than in the isotropic scattering case.
We note that for $\omega\tau = 60$, ${\mathrm v}_F/s = 84$, the threshold value for
ZRS at $\omega_c/\omega > 1$ is ${\mathcal E} \simeq 0.003$.
In Fig.~\ref{fig:longshortcomp} we clearly demarcate the values of
$\omega_c/\omega$  where our analytic formulae are valid ($qR_c/\wct \ll 1$).  

\section{Discussion}
\label{sec:discuss}

\subsection{Combination of classical and quantum effects}
\label{sec:classquant}
At low frequencies ($\omega \ll \omega_c$) we can use the approach in 
Sec.~\ref{sec:qenergybalance} to obtain an analytic expression for the quantum 
resistivity correction, starting with the low frequency 
limit of the balance equation for
vanishing dc field.  This calculation was already considered in 
Ref. \onlinecite{Robinson}, and we quote the result here for completeness.  We find
a non-vanishing addition to both diagonal components of the conductivity
(and, therefore, also of the resistivity) even when $k_{B}T\gg \hbar \omega
_{c}$. This generates isotropic magneto-oscillations for short-range disorder
\begin{equation}
\dfrac{\delta ^{q}\rho _{\alpha \alpha }}{\rho_0}=\dfrac{\delta
^{q}\sigma _{\alpha \alpha }}{\sigma_0}=-\frac{2\tau _{in}}{\tau }%
\left\vert \frac{4\pi \Gamma \epsilon _{F}}{\hbar \omega _{c}}\right\vert
^{2}\mathcal{E}^{2}J_{0}^{2}\left( qR_{c}\right) ,
\end{equation}%
(where $\sigma_0 = 1/\rho_0$)
in addition to the anisotropic classical commensurability effect.
Together, they yield the result:
\begin{eqnarray}
\frac{\delta \rho _{xx}}{\rho _0} &\approx &\frac{\delta \sigma _{yy}}{%
\sigma _0}=2J_{0}^{2}\left( qR_{c}\right) \mathcal{E}^{2}\left[ \frac{%
\mathrm{v}_{F}^{2}}{s^{2}}-\frac{\tau _{in}}{\tau }\left( 2\pi \Gamma \nu
\right) ^{2}\right] ,  \notag \\
\frac{\delta \rho _{yy}}{\rho _0} &\approx &\frac{\delta \sigma _{xx}}{%
\sigma _0}=-2J_{0}^{2}\left( qR_{c}\right) \mathcal{E}^{2}\frac{%
\tau _{in}}{\tau }\left( 2\pi \Gamma \nu \right) ^{2},  \label{FullResult}
\end{eqnarray}
valid when 
$\tau ^{-1}\lesssim \omega \ll \omega _{c}$,
where $\nu =2\epsilon _{F}/\hbar \omega _{c}$ is
the filling factor.  For a long-range potential, the oscillations are 
very similar, for ${\tau^*}^{-1} \lesssim \omega \ll \omega_c$, with $\tau^*$ replacing $\tau$ to give 

\begin{eqnarray}
\frac{\delta \rho _{xx}}{\rho _0} &\approx &\frac{\delta \sigma _{yy}}{%
\sigma _0}=2J_{0}^{2}\left( qR_{c}\right) \mathcal{E}^{2}\left[ \frac{%
\mathrm{v}_{F}^{2}}{s^{2}}\frac{\tau}{\tau^*}  
-\frac{\tau _{in}}{\tau^* }\left( 2\pi \Gamma \nu
\right) ^{2}\right] ,  \notag \\
\frac{\delta \rho _{yy}}{\rho _0} &\approx &\frac{\delta \sigma _{xx}}{%
\sigma _0}=-2J_{0}^{2}\left( qR_{c}\right) \mathcal{E}^{2} \frac{%
\tau _{in}}{\tau^* }\left( 2\pi \Gamma \nu \right) ^{2} . \label{FullResultstar}
\end{eqnarray}
The actual magnetoresistance
trace observed in an experiment will have both anisotropic classical and
isotropic quantum contributions.  The parameter that controls the overall
behaviour for both isotropic scattering and small-angle scattering is

\begin{equation}
\eta
\equiv \left( 2\pi \Gamma \nu s/\mathrm{v}_{F}\right) \sqrt{\tau _{in}/\tau } .
\end{equation}
 We can use $\eta$ to identify three different magnetoresistance
regimes.  If $\eta < 1$, the observed change in the resistivity is in
the direction of SAW propagation.  If $\eta \simeq 1$, then the
observed change in the resistivity is in the direction perpendicular
to the SAW wavevector, and if $\eta \gg 1$, then the resistance
correction is isotropic and negative.  For the parameter values used
in Figs.~\ref{fig:quant} and \ref{fig:longshortcomp}, the parameter $\eta$
is large compared to one, such that  the quantum contribution dominates the
classical corrections.

The quantum contribution to the resistivity that we discuss in Sec.~\ref{sec:quantumKE}
is not the only possible contribution to the resistivity that can arise from
quantum effects neglected in a classical calculation. 
A static periodic potential affects the
equilibrium density of states, which can lead
to corrections to $\rho_{yy}$ as well as to $\rho_{xx}$.\cite{Gerhardts,Gerhardts2}  
 These corrections
were ignored in our calculations, but at least in the static case
appear to be at most the order of magnitude of the classical contribution.
Hence the quantum effect we discuss here should dominate that quantum correction for
$\eta > 1$, provided $\tau_{in} \gg \tau$ and $\nu > {\mathrm v}_F/2\pi s$.

\subsection{Experimental Implications}
There are four situations we have discussed in this paper: the combinations of either
short- or long-range disorder and quantum and classical corrections to the magnetoresistance
due to SAWs.  We first focus on the case of isotropic scattering.  The most
pronounced geometric and temporal oscillations in the magnetoresistance are in 
the region $\omega_c/\omega \lesssim 1$ regardless of whether the oscillations are
quantum or classical in origin.  The classical effects dominate for short
inelastic scattering times and lead to anisotropic, positive magnetoresistance
corrections, whilst quantum effects dominate for large inelastic
scattering times, and can lead to ZRS which are modulated by geometric commensurability 
oscillations.  For small-angle scattering, geometric commensurability oscillations in 
the regime $\omega_c/\omega \lesssim 1$ are very strongly damped in both the classical and 
quantum cases, unless $(qR_c)^2/\wct \lesssim 1$ in this region.  This condition
is not met in current high quality 2DEGs in which small-angle scattering dominates.
However, it is currently possible to achieve $(qR_c)^2/\wct \lesssim 1$ for 
$\omega_c/\omega > 1$, and these geometric commensurability oscillations should be
observable either in the quantum or classical regimes when $\omega_c/\omega > 1$.  
Similarly to isotropic
scattering, quantum effects dominate for large inelastic scattering times (compared to 
the elastic scattering time) and classical effects are more important for short
inelastic scattering times.  It appears that the ZRS at low frequencies 
we predicted in Ref.~\onlinecite{Robinson} for isotropic scattering, are hard to achieve
if the scattering is isotropic, but should be much more readily achievable in samples
where small-angle scattering dominates, which is the experimentally relevant situation.

One theoretical prediction for ZRS is that they require 
inhomogeneous current flow in a 2DEG, 
\cite{Andreev} which appears to have 
recently been observed for microwaves.\cite{Willett2}
A large enough
SAW-induced change $\left\vert \delta \sigma _{xx}\right\vert >\sigma _{0}$
resulting in negative local conductivity would also require the formation of electric
field/Hall current domains. Since the anisotropy in Eq.~(\ref{FullResult})
suggests that such conditions can be achieved most easily in the
conductivity component along the SAW wavevector, we expect that domains
would form with current flowing perpendicular to the direction of SAW
propagation, and their stability would depend on the sample geometry. 
For a SAW with the wavevector directed across the axis of a Hall bar, 
current domains can be stabilized by ending in ohmic contacts. 
For a wave propagating (or standing) along the Hall bar, 
current domains would have to orient across the bar direction 
and terminate at the sample edges (destabilizing them), 
leading to a finite resistance. 
Finally, the anisotropy would not support a zero-conductance 
regime in a Corbino geometry.

In addition to the parameter ranges that are optimal for observing SAW-induced
ZRS, there are several other experimental issues we would like to mention.  
Firstly there is the observation by Kukushkin {\it et al.} \cite{KuKuMPL}
that microwave irradiation of high quality 2DEGs can lead to SAW generation.  
Thus one might want to consider the effects of microwaves and SAW simultaneously 
-- such a scheme might also allow for probing a 2DEG 
at frequencies other than $qs$ for a given SAW wavenumber.  However, 
this calculation is beyond the scope of the present work.\cite{footnote2}  

If increases in SAW frequency and  2DEG mobility allow
$\omega\tau \gg (qR_c)^2$ for $\omega/\omega_c \geq 1$, the geometric 
oscillatory structure similar to 
that observed for isotropic scattering i.e. peaks when $\omega/\omega_c$ is an integer 
should be visible in both classical and quantum contributions to magnetoresistance.
This is currently not yet observable for samples in which small-angle scattering
dominates (as appears to be the case for the samples used in Refs.~\onlinecite{Mani1,Zudov2}
where ZRS were observed) because $\omega_c\tau^* \gg 1$ when $\omega/\omega_c \gtrsim 1$.
Either samples in which isotropic scattering dominates are required to see these effects,
or higher quality samples in which $(qR_c)^2/\wct \gg 1$ when $\omega/\omega_c > 1$ are
required.  In terms of trying to improve the possibility of observing interesting
SAW induced magnetoresistance effects in the frequency range $\omega \gtrsim \omega_c$
in samples where small-angle scattering dominates, the following considerations
may be helpful. Since ${\mathrm v}_F/s \propto n_e^\frac{1}{2}$ and $\omega\tau \propto 
\omega\mu$ (and it is found that $\mu \propto n_e^{0.7}$ in the highest quality 
GaAs/AlGaAs 2DEGs \cite{Pfeiffer}), the best hope for observing effects 
with $\omega\tau \gg {\mathrm v}_F/s$ appears to be by achieving higher SAW frequencies.

\subsection{Conclusions}

In conclusion, we have demonstrated a new class of magnetoresistance
oscillations caused in a 2DEG by SAWs. We have shown that at $\omega \ll \omega_c$ 
the effect consists of contributions with competing signs: (i) a classical geometric
commensurability effect analogous to that found in static systems with
positive sign, and (ii) a quantum correction, with negative sign for either
isotropic or small-angle scattering. The latter
result suggests that SAW propagation through a high mobility electron gas
may generate a sequence of zero-resistance states (ZRS) linked to the maxima
of $J_{0}^{2}\left( qR_{c}\right) $ for strong enough SAW fields.  We find that
in the region $\omega \ll \omega_c$, SAW-induced ZRS states are much more
likely to be observed in samples for which small-angle scattering dominates
isotropic scattering.  Whilst
this prediction concerns the low-frequency domain $\omega \lesssim \omega
_{c}$, such ZRS would be formed via the same mechanism \cite{Andreev,Aleiner}
as the microwave-induced ZRS at $\omega \gtrsim \omega _{c}$.  
In the regime $\omega \gtrsim \omega_c$, we find that there are geometric
oscillations superposed on ZRS if there is isotropic scattering.  If there
is small-angle scattering, such oscillations are unlikely to be seen in 
2DEGs at the present time.  Hence the optimal parameter region to search for
SAW induced ZRS in 2DEGs with long-range disorder 
which show geometric modulation is for $\omega < \omega_c$ 
and $(qR_c)^2/\wct < 1$.

We thank I. Aleiner, D. Khmelnitskii, and A. Mirlin for discussions. This work was
funded by EPSRC grants GR/R99027, GR/R17140, and the Lancaster Portfolio
Partnership. It progressed during the
Workshop ``Quantum transport and correlations in mesoscopic systems and Quantum Hall
Effect'' at the Max-Planck-Institut PKS in Dresden, 
the Workshop on ``Quantum Systems out of Equilibrium'' at the Abdus Salam ICTP in Trieste,
and V. F.'s research visit to ICTP.


\begin{references}
\bibitem{footnote1} 
 A similar structure to $\tau_{N q}$ was also
found in Ref.~\onlinecite{Rudin} whilst studying 
density of states correlations in a 2DEG in a classical magnetic field.
\end{references}
\end{document}